\documentclass[11pt]{article}
\pdfoutput=1   	
\usepackage[margin=1in]{geometry}
\usepackage{graphicx}				
\usepackage{amssymb}
\usepackage{amsmath}
\usepackage{amsthm}
\usepackage{color}
\usepackage{changepage}
\usepackage{float}
\usepackage{cite}

\usepackage{caption}
\usepackage{subcaption}

\newtheorem{defn}{Definition}

\usepackage{titlesec}

\titlespacing\section{0pt}{14pt plus 4pt minus 2pt}{0pt plus 2pt minus 2pt}
\titlespacing\subsection{0pt}{16pt plus 4pt minus 2pt}{0pt plus 2pt minus 2pt}
\titlespacing\subsubsection{0pt}{14pt plus 4pt minus 2pt}{0pt plus 2pt minus 2pt}

\titleformat*{\section}{\Large\bfseries}
\titleformat*{\subsection}{\Large\bfseries}
\titleformat*{\subsubsection}{\large\bfseries\itshape}

\title{Selecting Between-Sample RNA-Seq Normalization Methods from the Perspective of their Assumptions}
\author{Ciaran Evans, Johanna Hardin, and Daniel Stoebel}

\begin{document}
\maketitle

\begin{small}
\noindent \textbf{Ciaran Evans} is a PhD student in statistics at Carnegie Mellon University. He is interested in applications of statistics to high-throughput genetic data.\\
\textbf{Johanna Hardin} is Professor of Mathematics at Pomona College. She works on different types of analyses of high-throughput genetic data that don't conform to the usual assumptions needed for statistical analyses.\\
\textbf{Daniel Stoebel} is Associate Professor of Biology at Harvey Mudd College. He works on the molecular biology and evolution of bacterial regulatory networks, particularly the \textit{E. coli} global transcriptional regulator RpoS.
\end{small}

\section*{ABSTRACT}
RNA-Seq is a widely-used method for studying the behavior of genes under different biological conditions. An essential step in an RNA-Seq study is normalization, in which raw data are adjusted to account for factors that prevent direct comparison of expression measures. Errors in normalization can have a significant impact on downstream analysis, such as inflated false positives in differential expression analysis. An under-emphasized feature of normalization is the assumptions upon which the methods rely and how the validity of these assumptions can have a substantial impact on the performance of the methods. In this paper, we explain how assumptions provide the link between raw RNA-Seq read counts and meaningful measures of gene expression. We examine normalization methods from the perspective of their assumptions, as an understanding of methodological assumptions is necessary for choosing methods appropriate for the data at hand. Furthermore, we discuss why normalization methods perform poorly when their assumptions are violated and how this causes problems in subsequent analysis. To analyze a biological experiment, researchers must select a normalization method with assumptions that are met and that produces a meaningful measure of expression for the given experiment.
\\
\emph{\textbf{Key words:} RNA-Seq; normalization; assumptions; differential expression; spike-in control; transcriptome size}

\section*{INTRODUCTION}
The introduction of microarrays provided the ability to study many genes in an organism under different biological conditions, with a dramatic reduction in expense and time from previous methods \cite{shendure2008beginning}. More recently, high-throughput sequencing has become an affordable and effective way of examining gene behavior and has been applied to a wide range of biological studies. For example, very specific questions about transcriptomes and splicing can now be addressed \cite{oshlack2010rna}, and the study of techniques for the analysis of high-throughput sequencing data continues to be a hot topic, involving researchers from biology, statistics, and computer science.

High-throughput sequencing with RNA, commonly referred to as RNA-Seq, involves mapping sequenced fragments of cDNA. In RNA-Seq, the RNA is fragmented and then reversed transcribed to cDNA (or reverse transcribed then fragmented). These fragments are then sequenced, producing \textit{reads} which are aligned back to a pre-sequenced reference genome or transcriptome \cite{oshlack2010rna, wang2009rna, auer2011differential}, or in some cases assembled without the reference \cite{wang2009rna}. The number of reads mapped to a gene is used to quantify its expression.

To convert raw read counts into informative measures of gene expression, \textit{normalization} is needed to account for factors that affect the number of reads mapped to a gene, like length \cite{oshlack2009length}, GC-content \cite{risso2011gc}, and sequencing depth \cite{robinson2010scaling}. Length and GC-content are \textit{within-sample} effects, meaning that they affect the comparison of read counts between different genes in a sample. Sequencing depth, on the other hand, is a \textit{between-sample} effect that alters the comparison of read counts between the same gene in different samples. Here we focus on between-sample normalization, which is needed to account for \textit{technical effects} (differences not due to the biological conditions of interest) that prevent read count data from accurately reflecting differences in expression \cite{robinson2010scaling}. In RNA-Seq, a cDNA library is constructed and then a portion of the molecules are sequenced to produce reads \cite{mcintyre2011technical}. Experimental variability, such as variability in the total number of molecules sequenced, can lead to different total read counts in different samples; this is referred to as differences in \textit{sequencing depth}, and the total number of reads in a sample is the \textit{library size} of that sample \cite{dillies2013comprehensive}. When one sample has more reads than another, non-differentially expressed genes will tend to have higher read counts in that sample \cite{robinson2010scaling} and so a correction is necessary. For applications requiring both between-sample and within-sample normalization, performing both types of normalization may be necessary; for example, Risso \textit{et al.} recommend using within-sample GC-content normalization in combination with between-sample normalization \cite{risso2011gc}.

Many normalization schemes have been proposed to account for between-sample effects in RNA-Seq data \cite{dillies2013comprehensive}, and several attempts have been made to determine the best strategy \cite{bullard2010evaluation, dillies2013comprehensive, kadota2012normalization, li2015comparing, lin2016comparison, maza2013comparison, rapaport2013comprehensive, zyprych2015impact}. However, very little attention has been paid to the assumptions upon which the different normalization methods rely. Several authors have identified situations in which a few highly expressed genes make up a large proportion of the total reads \cite{bullard2010evaluation, dillies2013comprehensive, lin2016comparison}, which could result in differences in distribution of read counts among genes. Others have found cases in which most or all genes are up-regulated in one condition \cite{athanasiadou2016growth, hu2014nucleosome, lin2012transcriptional, nie2012cMyc}. These situations, especially a global shift in expression, violate assumptions of many commonly-used methods and so result in errors in downstream analysis. Furthermore, biological experiments in which assumptions are unwittingly violated may mean that there are flaws in comparisons of normalization methods and in the conclusions drawn from these experiments.  As we have evidence of violated assumptions in some biological experiments, but not the extent to which assumptions are violated in others, it has been suggested that many prior conclusions are incorrect and a reanalysis of published results is necessary \cite{chen2015overlooked}.

The goal of this paper is to present normalization methods in the context of their assumptions, and to evaluate the effect and importance of assumptions on the performance of different normalization methods. We believe that a focus on assumptions can aid in evaluating different methods, and in choosing an appropriate method given knowledge of which assumptions are reasonable to make for the experiment at hand. With this in mind, we group between-sample methods by the assumptions they rely on and their strategy for normalization. We explain the reason the assumptions are necessary and the result of using a method when the assumptions do not hold. Finally, we examine previous research that aims to determine \textit{which} normalization method is better from the perspective of \textit{why} some methods perform better than other in specific situations.

\section*{GENE EXPRESSION AND NORMALIZATION}
The goal of normalization is for differences in normalized read counts to represent differences in true expression. Normalization is correct when the relationship between normalized read counts is correct. This essential purpose of normalization clearly depends on what we mean by ``true expression." In a simple view of gene expression, DNA is transcribed to RNA, which is then translated into protein. As discussed by Pachter \cite{pachter2011models}, RNA-Seq read counts do not measure final production of protein. Rather, sequencing technology quantifies the intermediate of gene expression: RNA, often specifically mRNA.  Given that the actual product of gene expression is never measured,  we consider the true expression of a gene to be the amount of mRNA/cell it produces.

This appears to be the definition commonly used in previous work, as prior research considers a gene to be \textit{differentially expressed} (DE) across different biological conditions if there is a difference in the amount of mRNA/cell it produces under these conditions. For example, authors discussing a global shift in expression talk about a global change in the absolute amount of RNA from a fixed number of cells \cite{loven2012revisiting}. In this paper we view expression and differential expression in terms of absolute quantities of mRNA/cell, and keeping this perspective in mind is important for understanding our discussion of normalization methods and their assumptions. However, it is important to note that other definitions of expression and differential expression are possible \cite{coate2015variation}, and beginning with a different definition may change which methods are appropriate for a given RNA-Seq experiment. For example, for certain biological experiments one might be interested in detecting differences in mRNA/transcriptome (that is, a gene's proportion of mRNA out of all mRNA transcribed) rather than mRNA/cell \cite{coate2015variation}.

Considering differentially expressed genes is very helpful for understanding normalization. As stated above, correct normalization will result in correct relationships between normalized read counts. In terms of differential expression, this means that non-DE genes should on average have the same normalized read counts across conditions, while DE genes should have normalized read counts whose differences (ratios) across conditions represent the true differences (ratios) in mRNA/cell. As with microarrays, a common use of RNA-Seq is to investigate the differential expression of an organism's genes under different biological conditions \cite{oshlack2010rna}, but normalization is needed in any RNA-Seq study where the relationship between normalized read counts must be correct, not just in differential expression analyses. In this paper, for simplicity we restrict our examples to the most basic case of two biological conditions, which will generally be referred to as A and B. Our results, however, hold for any number of conditions.

Gene expression is measured with RNA-Seq using the number of reads aligned to each gene under each biological condition \cite{wang2009rna}. However, a naive comparison of read counts for a given gene under the different conditions is problematic for two reasons. First, the number of reads aligned to a given gene in a given sample can be considered a random variable \cite{anders2010differential}, and so read count comparisons must take into account the variability of these random variables; an observed difference in count could simply be due to random chance.  Second, the total number of reads can vary across samples \cite{oshlack2010rna}, and so a large difference in a gene's read count between different conditions may simply be the result of differential coverage, rather than of differential expression. It is the second problem that necessitates normalization of read counts before differential expression analysis can be performed \cite{oshlack2010rna, auer2011differential}.

Normalization is an essential step in an RNA-Seq analysis, in which the read count matrix is transformed to allow for meaningful comparisons of counts across samples. With the advent of RNA-Seq technology it was initially believed that normalization would not be necessary \cite{wang2009rna}, but normalization has been found to be indispensable for correct analysis of RNA-Seq data.  Indeed, Bullard \textit{et al.} \cite{bullard2010evaluation} found that the normalization procedure used in a differential expression pipeline had the largest impact on the results of the analysis, even more than the choice of test statistic used in hypothesis tests for differential expression.

\begin{figure}
\centering
\includegraphics[scale=0.48]{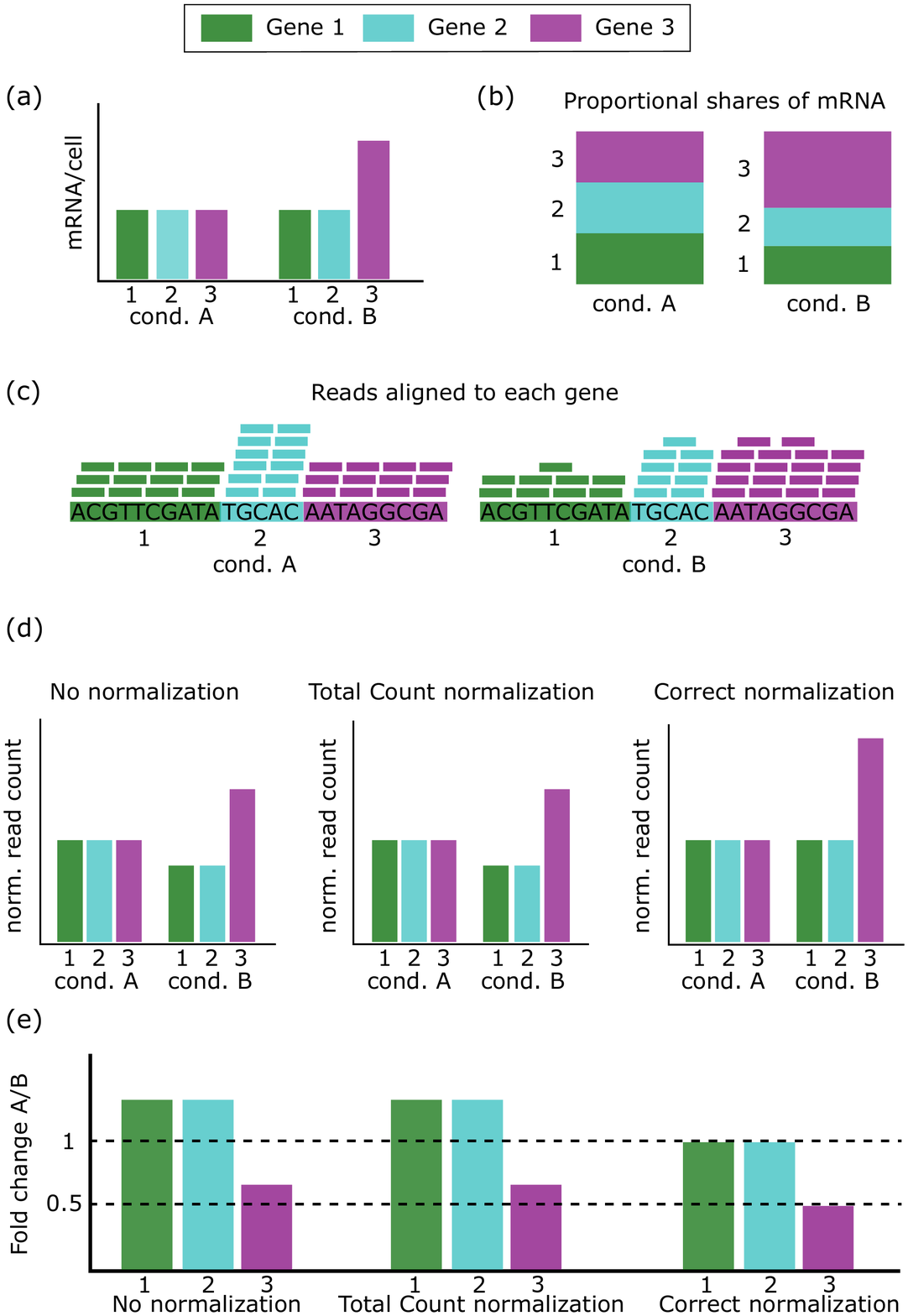}
\caption{
\small{
\textbf{One highly expressed gene}. An experiment is performed with conditions A and B to compare expression for the three genes (1, 2, and 3). \textbf{(a)} Gene 3 is two-fold up-regulated under condition B, while the other genes are not differentially expressed; the quantity of mRNA/cell (in bp) is the same for Genes 1 and 2, but is twice as high for Gene 3 under condition B. \textbf{(b)} Because of the change in expression of Gene 3, the shares of mRNA in the cell are different between conditions.  Under condition A each gene gets 1/3, whereas under condition B Gene 3 gets 1/2 while the other two get 1/4. \textbf{(c)} Differences in shares of mRNA are reflected in the shares of reads. Each sample has the same total number of reads, but the distribution is different between the conditions, matching the distribution of mRNA in (b). \textbf{(d)} When no normalization is performed, there are apparent differences in read counts for all three genes. Total Count normalization produces the exact same result as no normalization at all, since the total read count for each sample is the same. In truth there is no difference in expression for Genes 1 and 2, and the relative count for Gene 3 should be higher than found by no normalization or Total Count normalization. Correct normalization, therefore, makes the read counts of the non-differentially expressed genes equivalent, which also makes the relative expression of Gene 3 correct. \textbf{(e)} No normalization and Total Count normalization fail to equilibrate the read counts of the non-DE genes, resulting in each gene appearing differentially expressed, and the truly DE gene (Gene 3) having the wrong fold change. Correct normalization reveals no difference in expression for the non-DE genes and the correct fold change for Gene 3.}}
\label{fig: oneOverExpressed}
\end{figure}

Another reason normalization is required is that the proportion of mRNA corresponding to a given gene may change across biological conditions. In the sample of molecules sequenced, the number of molecules (and so by extension the number of reads) corresponding to a given gene is tied to that gene's share of the population of molecules available for sequencing. Hence, when there are a few genes that are highly expressed in only one of the conditions, the few genes will make up a greater share of the total molecules and so a smaller fraction of the reads will be left for the other genes \cite{robinson2010scaling}. This can cause the false appearance of differential expression for the non-DE genes, and normalization is needed to account for this difference. A visualization of such a situation is presented in Figure \ref{fig: oneOverExpressed}. Of the three genes, one is up-regulated while the other two are non-DE (Figure \ref{fig: oneOverExpressed}(a)). The one highly expressed gene leads to differences in shares of the proportion of mRNA for each gene (Figure \ref{fig: oneOverExpressed}(b)) which in turn causes differences in the share of reads aligned to each gene, even if the total number of reads is the same in each condition (Figure \ref{fig: oneOverExpressed}(c)). If the differences in read share are not corrected by normalization (Figure \ref{fig: oneOverExpressed}(d)) then the apparent fold change for every gene will be wrong (Figure \ref{fig: oneOverExpressed}(e)). Correct normalization, on the other hand, equilibrates the read counts for the two non-DE genes (Figure \ref{fig: oneOverExpressed}(d)) and thereby leads to accurate observed fold changes (Figure \ref{fig: oneOverExpressed}(e)).

\begin{figure}
\centering
\includegraphics[scale=0.48]{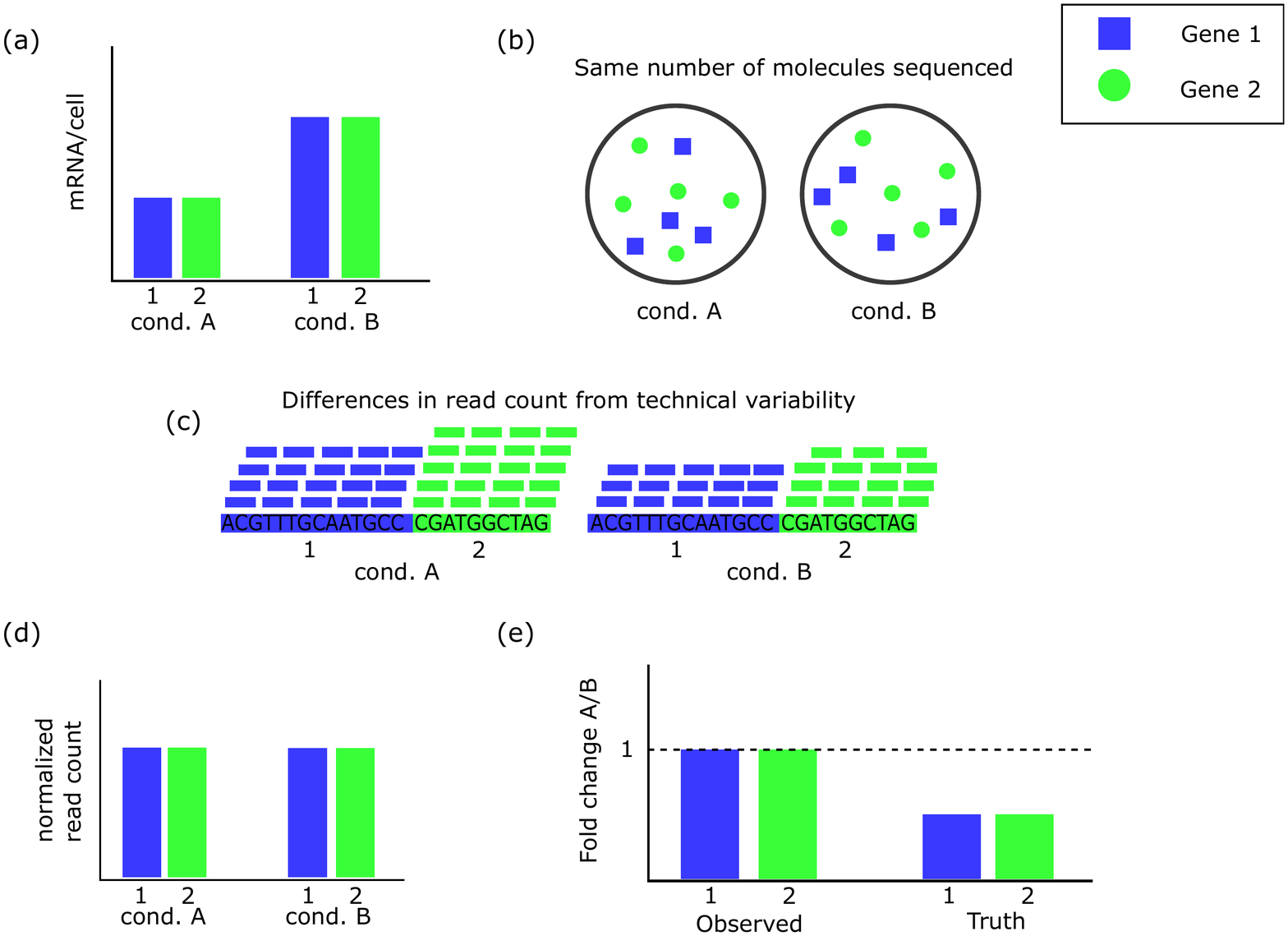}
\caption{\small{\textbf{Global shift in expression}. There are two genes, and an experiment is performed to compare expression between condition A and condition B. \textbf{(a)} There is global up-regulation under condition B vs. condition A, with both genes having twice the expression under condition B. Within each condition, the two genes produce the same amount of mRNA/cell (measured in bp). \textbf{(b)} In the RNA-Seq experiment, the same number of molecules are sequenced from each of the two samples. Proportionally, the mRNA composition is the same under each condition, so the composition of molecules sequenced is also the same. Within each condition, the two genes produce the same amount of mRNA (in bp) but Gene 2 is 4/5 the length of Gene 1, so must produce 5/4 the number of molecules that Gene 1 does. \textbf{(c)} Sequenced reads are aligned to the reference genome and mapped to each gene. The distribution of reads is the same in each sample, but by chance the sample for condition A happens to have more reads in total. \textbf{(d)} Normalization is performed, which removes the differences in read count from technical variability, so the read count for each gene is the same across conditions.  \textbf{(e)} Because the normalized read counts are the same, the observed fold change for each gene is 1, indicating no differential expression. However, genes are really twice as expressed under condition B and so in truth we should see half the expression when comparing A to B.}}
\label{fig: allUpReg}
\end{figure}

As normalization methods have developed, it has become clear that initial approaches fail in cases of a shift in expression for many or all genes \cite{loven2012revisiting}. In cases like Figure \ref{fig: oneOverExpressed}, a small number of highly expressed genes creates the appearance that non-DE are differentially expressed, but the false DE calls may be corrected by normalizing read counts so that the expression levels of non-DE genes are equivalent.  In contrast, in the case of a global shift in expression it may appear that differentially expressed genes are non-DE or that up-regulated genes are down-regulated \cite{loven2012revisiting}. An example is presented in Figure \ref{fig: allUpReg}. All genes are up-regulated two-fold under condition B (Figure \ref{fig: allUpReg}(a)), but roughly the same number of molecules are sequenced (Figure \ref{fig: allUpReg}(b)). This conceals the fact that one condition results in twice as much total expression, and the only differences in read counts between the two conditions is due to technical variability (e.g. sequencing depth) (Figure \ref{fig: allUpReg}(c)). Conventional normalization approaches account for the technical differences, resulting in the same normalized read counts under each condition (Figure \ref{fig: allUpReg}(d)). Conventional normalization fails to reflect the two-fold up-regulation under condition B, and examining the observed fold changes (Figure \ref{fig: allUpReg}(e)) it appears that neither gene is differentially expressed when in truth both are.  A further need for normalization is therefore in cases of global shifts in expression, in which it is necessary to take into account the differences in overall expression between conditions.

To address the variety of needs for normalization, a corresponding variety of normalization methods has been developed. To correctly normalize, each method requires one or more assumptions about the experiment and gene expression. Assumptions are necessary for converting read counts into meaningful measures of expression. In the following sections we organize normalization methods into groups of methods that rely on similar assumptions. 

\section*{NORMALIZATION METHODS AND ASSUMPTIONS}
Here we group normalization methods that have similar assumptions and approaches to normalization. Short descriptions of the methods are provided; more detailed information on the method specifics is available in the Supplementary Information.

Recall that for our purposes, a gene is differentially expressed across a set of conditions if that gene produces \textbf{different levels of mRNA/cell} under the different conditions. For a normalization method to work, the normalized read counts must be representative of the true mRNA/cell values. That is, if a gene produces twice as much mRNA/cell under condition A as under condition B, then the normalized read count for that gene should on average be twice as big under condition A as under condition B. However, RNA-Seq, on the other hand, initially produces relative measures of expression \cite{pachter2011models}. As shown in Figure \ref{fig: allUpReg}, the number of reads aligned to a given gene reflects the sequencing depth and that gene's share of the population of mRNA molecules. We shouldn't expect a gene with twice as much mRNA/cell to have twice the number of reads. To correctly normalize, then, we must make some assumptions so that measures of relative expression (raw read counts) can be translated into measures of absolute expression (normalized read counts).   Different groups of normalization methods discussed here take different approaches, and so require different assumptions to produce correctly normalized values. These assumptions often deal with the total amount of mRNA/cell or the amount of \textit{symmetry} in the differential expression.

\begin{figure}
\centering
\includegraphics[scale=0.48]{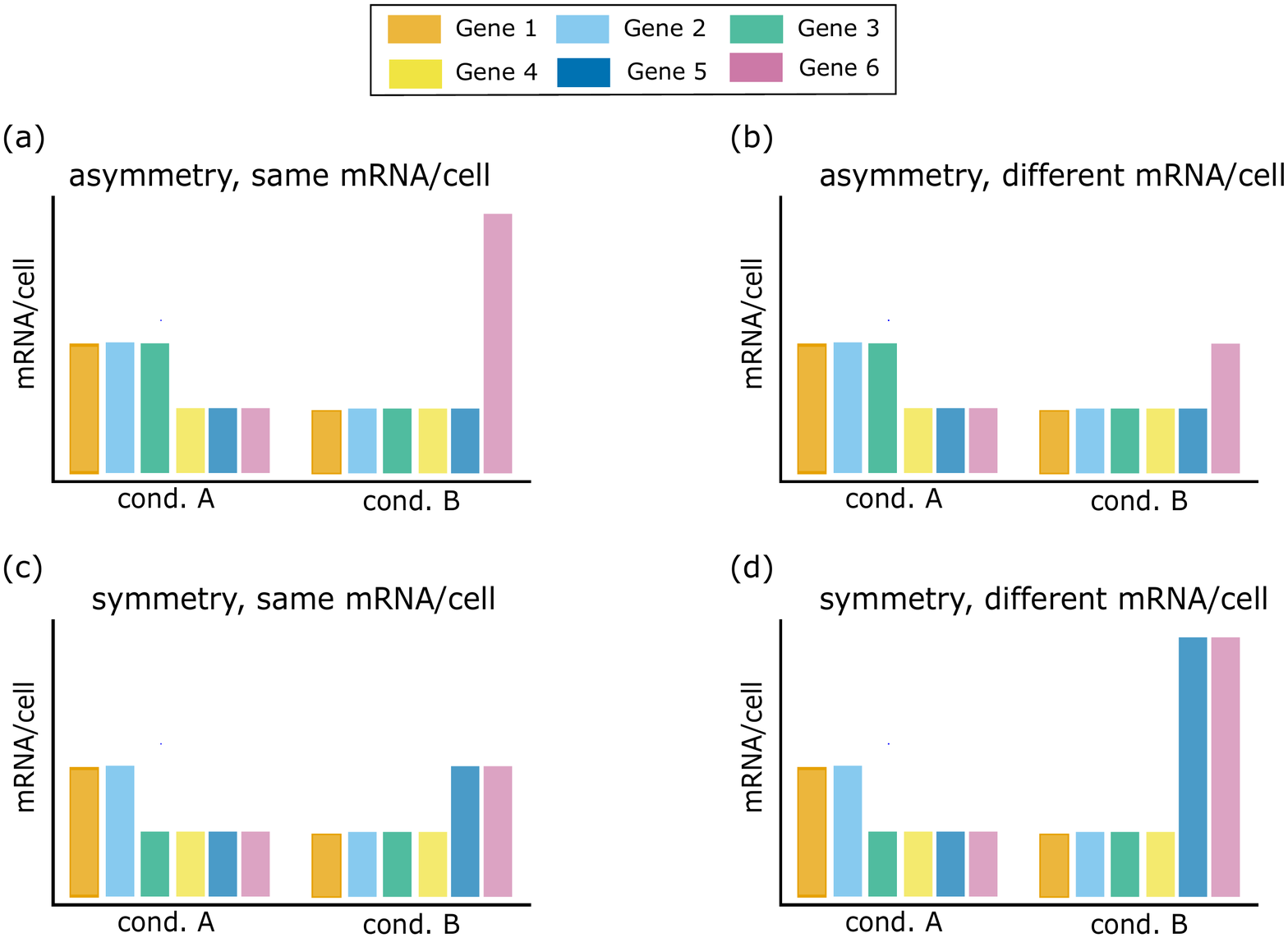}
\caption{\small{\textbf{Differential expression and (a)symmetry}.} There are six genes, and two experimental conditions. \textbf{(a)} Differential expression is asymmetric (3 up-regulated genes under condition A, 1 under condition B). The total mRNA/cell (summed over the six genes) is the same under both conditions. \textbf{(b)} Differential expression is asymmetric. The total mRNA/cell is different (less total mRNA/cell under condition B). \textbf{(c)} Differential expression is symmetric (2 up-regulated genes under each condition). The total mRNA/cell is the same under both conditions. \textbf{(d)} Differential expression is symmetric. The total mRNA/cell is different (more total mRNA/cell under condition B).}
\label{fig: DESymmQuad}
\end{figure}

We say that differential expression is \textit{symmetric} between two conditions when the number of genes up-regulated in each condition is equal. Figure \ref{fig: DESymmQuad} demonstrates the four possible combinations of symmetry/asymmetry and same/different total mRNA/cell. Figure \ref{fig: DESymmQuad} will be referenced to illustrate situations in which assumptions are and are not met.

\subsection*{Normalization by Library Size}
The normalization by library size aims to remove differences in sequencing depth simply by dividing by the total number of reads in each sample \cite{dillies2013comprehensive}.

\subsubsection*{Assumptions}
\textbf{Same total expression}: the amount of total expression is the same under the different experimental conditions. That is, each condition has the same amount of mRNA/cell. Figure \ref{fig: DESymmQuad}(a) and Figure \ref{fig: DESymmQuad}(c) show examples in which this assumption holds.

\subsubsection*{Methods}
\textbf{Total Count} normalization \cite{dillies2013comprehensive} divides each read count by the number of reads in its sample. The \textbf{RPKM} (reads per kilobase per million mapped reads) \cite{mortazavi2008mapping} and \textbf{FPKM} (fragments per kilobase per million mapped fragments) \cite{trapnell2010transcript} methods are essentially the same as Total Count normalization, but with the added component of accounting for gene length as well. \textbf{ERPKM} is a variant of RPKM that replaces gene length with an effective length \cite{li2015comparing}.

\subsubsection*{Motivation}
After dividing by library size, the normalized counts reflect the proportion of total mRNA/cell taken up by each gene. If the total mRNA/cell is the same across conditions, this proportion reflects absolute mRNA/cell for each gene.

\subsection*{Normalization by Distribution/Testing}
If technical effects are the same for DE and non-DE genes, then normalization could be done by equilibrating expression levels for non-DE genes. This set of methods attempts to capture information from non-DE genes. Normalization by distribution compares distributions (either of read counts or some function of read counts) across samples; normalization by testing attempts to detect a set of non-DE genes through hypothesis testing.

\subsubsection*{Assumptions}
\begin{enumerate}
\item \textbf{DE and non-DE genes behave the same:} technical effects are the same for DE and non-DE genes.
\item \textbf{Balanced expression:} there is roughly symmetric differential expression across conditions (same number of up-regulated and down-regulated genes). This assumption holds in Figure \ref{fig: DESymmQuad}(c) and Figure \ref{fig: DESymmQuad}(d). Normalization by testing can tolerate a larger difference in number of up- and down-regulated genes for higher proportions of DE than can normalization by distribution (see Figure \ref{fig: efdrPlots}).
\end{enumerate}

\subsubsection*{Methods}
\underline{Normalization by distribution}: \textbf{Quantile} normalization \cite{bolstad2003comparison} forces the distribution of the normalized data to be the same for each sample by replacing each quantile with the average (or median) of that quantile across all samples. Other methods do not force all quantiles to be the same, but instead focus on a specific quantile. \textbf{Upper Quartile} normalization \cite{bullard2010evaluation} divides each read count by the 75$^\text{th}$ percentile of the read counts in its sample. \textbf{Median} normalization \cite{dillies2013comprehensive} is essentially the same, but uses the median rather than the 75$^\text{th}$ percentile. The \textbf{DESeq} normalization method \cite{anders2010differential} finds the ratio of each read count to the geometric mean of all read counts for that gene across all samples (the denominator serving as a pseudo-reference sample \cite{anders2010differential}). The median of these ratios for a sample, called the \textit{size factor}, is used to scale that sample. This idea was expanded in the CuffDiff 2 software; \textbf{CuffDiff} normalization calculates \textit{internal} and \textit{external} size factors using the DESeq approach. The internal size factors are found for each sample by only considering other samples performed under the same biological condition when taking the geometric mean, while the external size factors are calculated after normalization by the internal size factors. The \textbf{TMM} (Trimmed Mean of the M-values) \cite{robinson2010scaling} approach is to choose a sample as a reference sample, then calculate fold changes and absolute expression levels relative to that sample. The genes are trimmed twice by these two values, to remove differentially expressed genes, then the trimmed mean of the fold changes is found for each sample. Read counts are scaled by this trimmed mean and the total count of their sample.  Note: the \textbf{edgeR} package \cite{robinson2010edgeR} uses TMM normalization, and so TMM could reasonably be called edgeR normalization instead. However, the name TMM seems to be more commonly used in the literature, and so we use it here.  \textbf{Median Ratio} normalization (MRN) \cite{maza2013comparison} is a method similar to TMM, with the goal of being more robust. In MRN, read counts are divided by the total count of their sample, then averaged across all samples in a condition for a given gene.  This produces an average count-normalized value for each gene and each condition, and the median of the ratios of these values between conditions is taken. The original counts are then normalized by this median and their library size.

\underline{Normalization by testing}: \textbf{PoissonSeq} \cite{li2012normalization} uses an iterative process that alternates between estimating a set of non-differentially expressed genes, and estimating the scaling factor for each sample using that set. Given estimates of the scaling factor, expected values for the read counts can be determined and non-DE genes are identified using a $\chi^2$ goodness-of-fit test. A similar iterative strategy is implemented by \textbf{DEGES} (Differentially Expressed Gene Elimination Strategy) \cite{kadota2012normalization}, which alternates between calculating scaling factors from a set of genes identified as non-DE, and estimating which genes are non-DE using differential expression hypothesis testing.

\subsubsection*{Motivation}
Non-DE genes should have, on average, the same normalized counts across conditions. Clearly, we want to normalize in order to equilibrate the non-DE genes. If technical effects impact non-DE genes and DE genes alike, then we can normalize all genes with the same normalization factor as the non-DE genes. So, we need to compare the non-DE genes; assuming balanced expression means we can estimate the differences in read counts between non-DE genes across samples.

\subsection*{Normalization by Controls}

\begin{figure}
\centering
\includegraphics[scale=0.48]{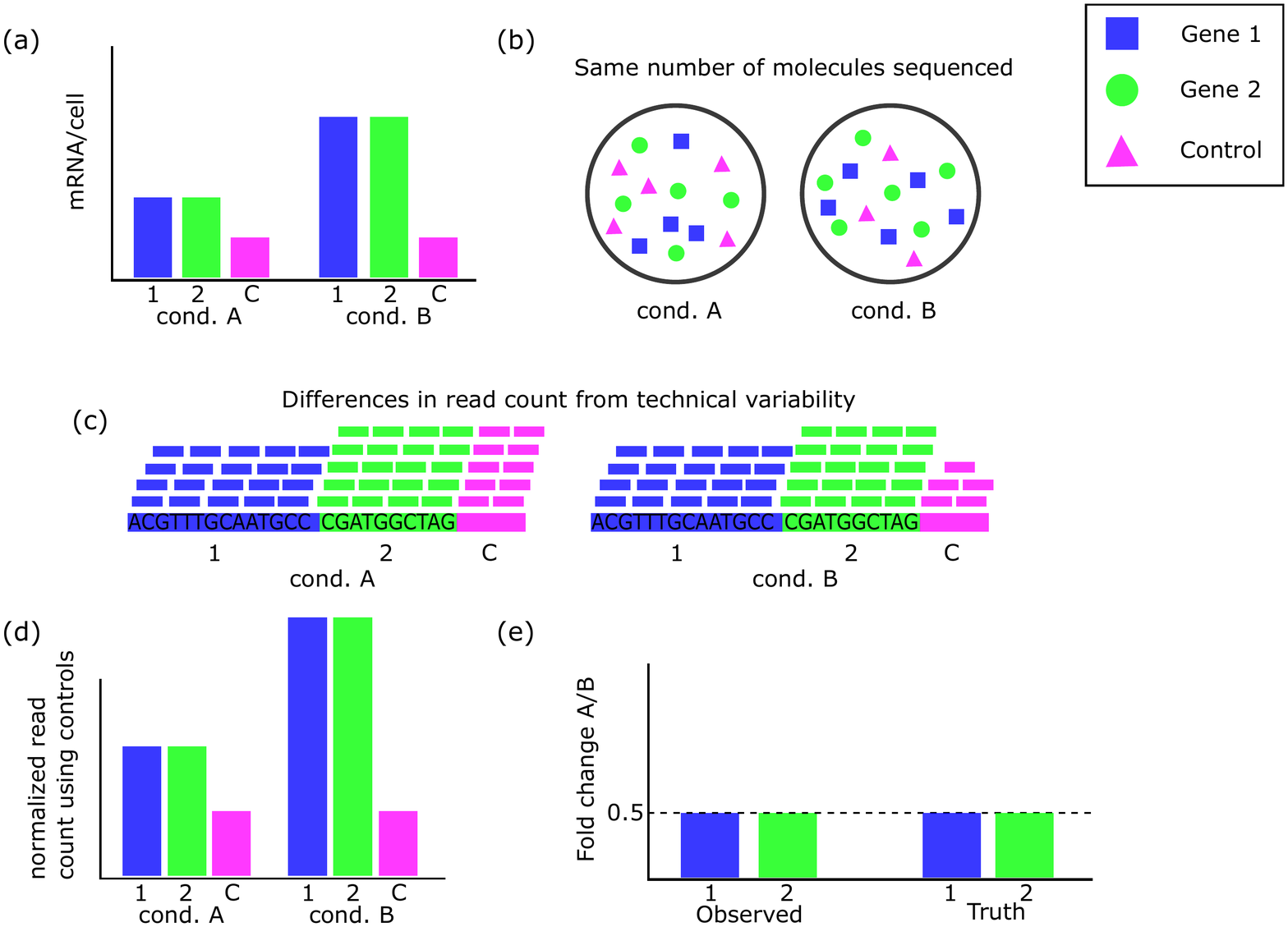}
\caption{ \small \textbf{Use of negative controls with shift in expression.} Two genes are investigated for differential expression between condition A and condition B. A negative control is used for normalization (could be a known non-DE gene or spike-in control). \textbf{(a)} Both non-control genes are up-regulated under condition B vs. condition A, having twice the expression under condition B. As a negative control, the control has the same expression under both conditions. \textbf{(b)} In the RNA-Seq experiment, the same number of molecules is sequenced from each sample. As the control has a smaller share of the mRNA in condition B, there are fewer control molecules in the sample for condition B. \textbf{(c)} Variability leads to differences in the total read count for the two samples. The share of the reads aligned to the control is the share of mRNA from the control. \textbf{(d)} The control should have the same expression in both conditions, so normalization is performed to equalize the normalized read count for the control, resulting in normalized read counts that reflect the correct mRNA/cell levels. \textbf{(e)} Because normalized counts correctly reflect mRNA/cell, the observed fold change agrees with the truth.}
\label{fig: simpleNegControl}
\end{figure}

Controls are needed for normalization when the assumptions of other methods are violated.  For example, Figure \ref{fig: allUpReg} demonstrates how a global shift in expression can go undetected. When controls are used, such as the negative controls illustrated in Figure \ref{fig: simpleNegControl}, then it is possible to correctly normalize by performing normalization on the controls. Since the controls are not affected by the biological conditions but the same amount of controls/cell are present in each condition (Figure \ref{fig: simpleNegControl}(a)) then different numbers of control molecules are sequenced (Figure \ref{fig: simpleNegControl}(b)). This leads to a share of the reads reflective of the share of mRNA for the control (Figure \ref{fig: simpleNegControl}(c)). By normalizing on the control, the correct levels of expression are seen (Figure \ref{fig: simpleNegControl}(d)) and so accurate fold changes are observed (Figure \ref{fig: simpleNegControl}(e)).

\subsubsection*{Assumptions}
\begin{enumerate}
\item \textbf{Existence of controls:} The controls needed for the experiment do in fact exist, and their expression behaves as expected (e.g., for negative controls they are non-DE under the conditions of the experiment).
\item \textbf{Controls behave like non-control genes:} The technical effects for the controls in some way reflect the technical effects for all the genes, so that the controls can be used for normalization.
\end{enumerate}

\subsubsection*{Methods}
\underline{Housekeeping genes}: \textbf{Housekeeping genes} (HG) are genes which play a role in the basic functions of a cell \cite{eisenberg2013housekeeping}, and so are believed to be non-DE under the biological conditions of interest \cite{bullard2010evaluation, eisenberg2013housekeeping}. HG normalization assumes that these genes are truly not differentially expressed, and furthermore that they are affected by technical effects the same way as DE genes. Normalization using housekeeping genes can either equalize the read count of the gene (if one housekeeping gene) \cite{bullard2010evaluation} or perform a conventional normalization procedure on a set of housekeeping genes \cite{dillies2013comprehensive}. 

\underline{Conventional normalization with spike-ins}: A set of synthetic \textbf{spike-in controls} is available through the ERCC \cite{jiang2011synthetic}, and these can be used instead of housekeeping genes. The use of spike-ins with conventional methods assumes that the spike-ins are not affected by the biological conditions under investigation, and that they have the same technical effects as the real genes \cite{risso2014normalization}. Conventional normalization methods, such as Upper Quartile, may be applied to the spike-ins \cite{risso2014normalization}, as with HG controls. The conventional normalization methods are applied only to the spike-ins, and then used to calculate normalization factors for all genes. One approach is proposed by Lov\'{e}n \textit{et al.} \cite{loven2012revisiting}, which uses cyclic loess normalization on the spike-ins (CLS). Spike-ins are added to RNA in proportion to the number of cells from which RNA is extracted. Then, cyclic loess normalization is performed on the RPKM values (more details can be found in the Supplementary Information). The loess curve is fit using only the spike-ins, but used to adjust all RPKM values so that the other genes are normalized with the spike-in information, which is not affected by differential expression.

\underline{Factor analysis of controls}: To address perceived problems with the use of spike-ins, \textbf{Remove Unwanted Variation (RUV)} \cite{risso2014normalization} uses factor analysis to remove factors of unwanted variation in RNA-Seq data. Using a set of negative control genes or samples, singular value decomposition is used to estimate a matrix for the factors of unwanted variation. Normalization to remove the factors of unwanted variation is then performed. It is divided into three sub-methods: RUVg, RUVs, and RUVr. The two assumptions listed above indicate slightly different things for the different sub-methods, and RUVr doesn't actually require controls (it is an adaptation of the RUV method to be used when controls are not available) \cite{risso2014normalization}. Here we list the meaning of the assumptions for each of the three sub-methods:
\begin{enumerate}
\item RUVg. Existence of controls: negative controls exist (non-DE across conditions). Controls behave like non-control genes: the factors of unwanted variation for the controls span the same space as the factors for the entire set of genes.
\item RUVs. Existence of controls: negative controls exist (non-DE across conditions) and there are also negative control samples (expression not related to biological condition). Controls behave like non-control genes: the factors of unwanted variation for the controls span the same space as the factors for the entire set of genes, and the factors of unwanted variation are not correlated with experimental condition.
\item RUVr. Does not require existence of controls. Assumes that factors of \textit{wanted} variation are known (i.e., the design matrix) and the factors of unwanted variation are not correlated with experimental condition.
\end{enumerate}

\subsubsection*{Motivation}
Controls should be non-DE across conditions and hence on average, normalized counts for the controls should be the same across conditions. If technical effects impact controls like they impact genes, then we can apply the adjustment for the controls to all genes. The reasoning for normalization by controls is similar to normalization by distribution/testing, but in the former it is assumed that an explicit set of controls is known, while in the latter we aim to capture the information from non-DE genes without knowing beforehand which genes are non-DE.

\section*{IMPORTANCE OF THE ASSUMPTIONS}

At first glance it makes sense that correcting for differences in sequencing depth can be done simply by library size normalization, which works if the total amount of mRNA in each cell is the same across experimental conditions. Then, a gene which produces the same amount of mRNA under each condition will produce the same \textit{proportion} of total mRNA in each condition. We thus expect the same proportion of reads to be aligned to that gene under each condition, and Total Count normalization gives us exactly the proportion of reads aligned to each gene.  Likewise, differences in expression correspond to differences in proportion of reads in the sample. However, differences in total mRNA/cell can lead to both failing to detect differentially expressed genes (Figure \ref{fig: allUpReg}) and incorrectly calling non-DE genes differentially expressed (Figure \ref{fig: oneOverExpressed}) when normalization by library size is performed in situations where total mRNA/cell is not constant.

On the other hand, normalization by distribution and by testing are impacted by differences in the number of up-regulated vs. down-regulated genes, but not by the relative amounts of mRNA/cell. The greater the disparity between the number of up-regulated genes and the number of down-regulated genes under a given condition, the higher the \textit{asymmetry} of the differential expression under that condition. Both Figure \ref{fig: oneOverExpressed} and Figure \ref{fig: allUpReg} show differences in total expression (mRNA/cell) between the two conditions, but there is much more asymmetry in Figure \ref{fig: allUpReg} (that is, 100$\%$ of the genes are up-regulated).  Accordingly, normalization by distribution and by testing can handle differences in mRNA/cell in the case of a few highly expressed genes (small asymmetry), but not a global shift in expression (large asymmetry). If there are only a few differentially expressed genes, these DE genes will not do much to change the estimated normalization factor. For example, the Upper Quartile normalization strategy compares the 75$^{\text{th}}$ percentile of read counts between samples.  If the 75$^{\text{th}}$ percentile of all the read counts is similar to the 75$^{\text{th}}$ percentile of the non-DE read counts, this is a reasonable approach. The normalization statistic for all genes will be similar to the normalization statistic for non-DE genes if there are only a few differentially expressed genes. The two statistics will also be similar when there is a small proportion of asymmetry. When differential expression is mostly symmetric, the values for differentially expressed genes should more or less balance out on either side of the statistic for non-DE genes, so that the statistic for all genes is close to the statistic for non-DE genes. A small proportion of asymmetry can allow distribution/testing methods to tolerate higher proportions of differential expression.

Knowledge of the assumptions made by each normalization method allows for good predictions of which biological experiments are suitable for each method. Normalization by library size should work well when total mRNA/cell is equivalent across conditions, regardless of the amount of asymmetry (Figure \ref{fig: DESymmQuad}(a) and Figure \ref{fig: DESymmQuad}(c)). On the other hand, normalization by distribution/testing should generally work well when there is symmetry, regardless of differences in mRNA/cell (Figure \ref{fig: DESymmQuad}(c) and Figure \ref{fig: DESymmQuad}(d)). When there is both asymmetry and different levels of total mRNA/cell (Figure \ref{fig: DESymmQuad}(b)), we expect both sets of methods to perform poorly.

\begin{figure}
\centering
\includegraphics[scale=0.65]{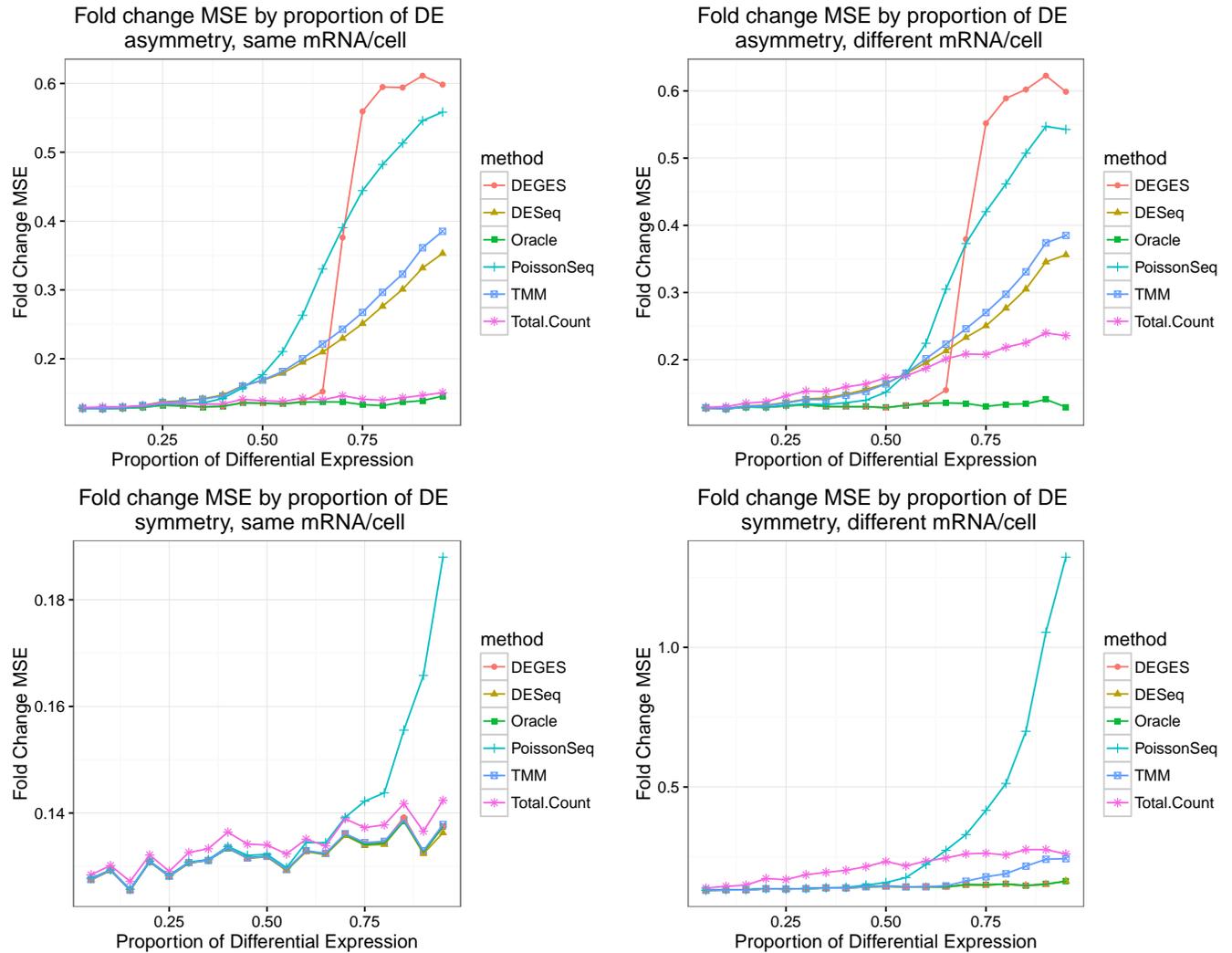}
\caption{\small \textbf{Impact of amount of asymmetry and amount of mRNA/cell on fold change estimates.} These plots show the average log fold-change MSE for non-DE genes of several methods. Simulated data is used, with varying proportions of differential expression (5$\%$ to 95$\%$). Genes simulated to be non-DE should have an observed log fold-change close to 0; the MSE is thus calculated by averaging the squared observed log fold-changes for each non-DE gene (treating the true log fold-change as 0). Because of variability in the generation of read count data, the observed log fold-change will in general not be exactly 0, so the Oracle normalization method (normalizing the data with the correct normalization factors given the simulation) serves as a baseline. Methods with MSEs that closely follow those of Oracle normalization are doing well. Asymmetric differential expression was simulated as 75$\%$ of the set of DE genes up-regulated in one condition and 25$\%$ up-regulated in the other. Under symmetric differential expression, 50$\%$ of DE genes are up-regulated in each condition. For simulations with the same mRNA/cell, non-DE genes had the same proportion of reads in each condition; simulations with different mRNA/cell resulted in non-DE genes having different shares of the reads in the different conditions.}
\label{fig: MSEPlots}
\end{figure}

\begin{figure}
\centering
\includegraphics[scale=0.65]{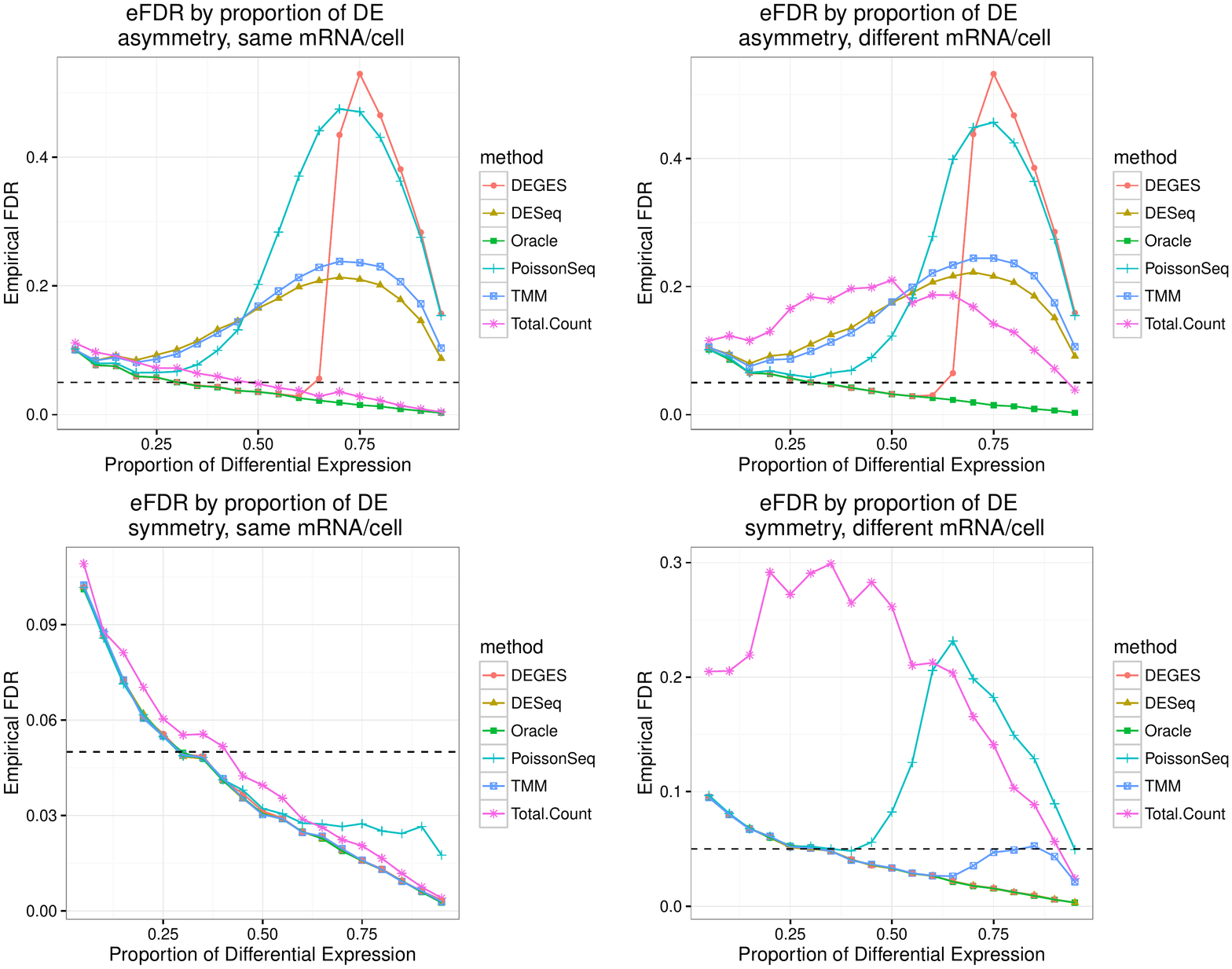}
\caption{ \small \textbf{Impact of amount of asymmetry and amount of mRNA/cell on error control.} These plots show the average empirical FDR of several methods on simulated data with varying proportions of differential expression (5$\%$ to 95$\%$). The simulations are performed with two conditions. Asymmetric differential expression was simulated as 75$\%$ of the set of DE genes up-regulated in one condition and $25\%$ up-regulated in the other. Under symmetric differential expression, 50$\%$ of DE genes are up-regulated in each condition. For simulations with the same mRNA/cell, non-DE genes had the same proportion of reads in each condition; simulations with different mRNA/cell resulted in non-DE genes having different shares of the reads in the different conditions. The black dashed line is at 0.05, the nominal FDR using the Benjamini-Hochberg adjustment. Deviations of the oracle value from the nominal value (starting above 0.05 and falling below as the proportion of DE increases) are a result of DESeq2 hypothesis testing and the conservativeness of Benjamini-Hochberg.}
\label{fig: efdrPlots}
\end{figure}

To demonstrate this, we examined the performance of several normalization methods on simulated data (Figure \ref{fig: MSEPlots} and Figure \ref{fig: efdrPlots}). For the simulations, we chose methods which were representative and generally perform well in the literature, as summarized in Table \ref{tab: litSummary} (except for Total Count normalization, as all normalization by library size methods perform poorly in the literature). We used Total Count, DESeq, TMM, PoissonSeq, DEGES, and finally Oracle normalization that uses the true normalization factor known from the simulation parameters. To measure how well the methods performed normalization, we used a method similar to Maza \textit{et al.} \cite{maza2013comparison} and calculated the mean squared error (MSE) of the log fold change for non-DE genes (Figure \ref{fig: MSEPlots}), comparing each observed log fold change to 0. As these genes are not differentially expressed, if normalization is performed correctly then the log fold change between samples of different conditions should be close to 0. Oracle normalization provides the baseline for the MSE under perfect normalization; methods which track closely with the Oracle are performing well.

Figure \ref{fig: MSEPlots} shows the results of the simulations, confirming that the methods perform as expected. Total Count normalization follows the Oracle closely when there is the same total mRNA/cell, but diverges quickly when there is different mRNA/cell. DESeq, TMM, and DEGES perform well when there is symmetry, for all proportions of differential expression. PoissonSeq does well under symmetry until too high a proportion of differential expression is reached, at which point it diverges. This is likely due to the fact that PoissonSeq normalization uses a set of genes of a fixed size for normalization; when the proportion of differential expression is too high, the set necessarily contains differentially expressed genes that skew the normalization estimate. When there is asymmetry, the normalization by distribution/testing methods can tolerate a small proportion of differential expression but eventually reach a break-down point.

The effects on downstream analysis of applying the different normalization methods are shown in Figure \ref{fig: efdrPlots}, which show empirical false discovery rate (eFDR) measures for each method after testing for differential expression (note: the downward trend in the Oracle eFDR is due to the use of the Benjamini-Hochberg procedure to control FDR, which is conservative and controls at a level directly related to the proportion of true null hypotheses, i.e. non-DE genes). When methods normalize correctly, as shown in Figure \ref{fig: MSEPlots}, the subsequent tests for differential expression are able to control the false discovery rate. However, when normalization fails and the observed fold changes depart sufficiently from the truth, the result is inflated false positives. This illustrates how heavily analysis relies on correct normalization, which in turn relies on assumptions. When the assumptions are violated, normalization fails (Figure \ref{fig: MSEPlots}) and as a result so does the downstream analysis (Figure \ref{fig: efdrPlots}).

\subsection*{Simulation Details}
To assess the downstream results of violating the assumptions of different normalization methods, simulations were run in which the average mean squared error (MSE), on non-DE log fold changes, and average empirical false discovery rate (eFDR) were computed for different proportions of differential expression (proportion of genes which are truly DE), amounts of asymmetry, and relative amounts of mRNA/cell. The code for the simulations and the plots of the results can be found at (https://github.com/ciaranlevans/rnaSeqAssumptions), and was adapted from the R code used in the simulations of Law \textit{et al.} \cite{law2014voom}.

Four sets of simulations were performed, one for each combination of asymmetry vs. symmetry and same mRNA/cell vs. different mRNA/cell. In each simulation, read count data was generated then normalized according to one of six different methods: DEGES, DESeq, Oracle (normalization with the true scaling factors, used for benchmarking other normalization methods), PoissonSeq, TMM, and Total Count. The normalization methods were selected to represent different types of normalization: by library size (Total Count), by distribution (DESeq and TMM), and by testing (PoissonSeq and DEGES). DESeq and TMM were chosen to represent normalization by distribution methods as they are widely studied and generally perform well relative to other methods (Table \ref{tab: litSummary}). Simulated RNA-Seq data was generated, then each normalization method was performed. After normalization, two normalized columns of the read count matrix (one from each condition) were compared to produce log fold changes for the non-DE genes. These observed log fold changes should be close to 0, so the MSE was calculated by averaging the squared log fold changes for the non-DE genes. Differential expression hypothesis testing was performed on the data for each normalization method. Testing was done separately from normalization, and was performed with the DESeq2 \cite{love2014moderated} package after normalization with each method (the data was not re-normalized with DESeq2). As in DESeq2, and as is common in differential expression studies, p-values were adjusted using the Benjamini-Hochberg procedure for FDR control \cite{benjamini1995controlling} (see Supplementary Information for more details). Using the adjusted p-values, and knowledge of which genes were simulated to be differentially expressed, the average eFDR (observed proportion of false discoveries out of all discoveries) was calculated across 50 repetitions.

Simulations begin by creating initial proportions of expression, representing the proportion of the total expression for each gene and each sample, with 1000 genes and 10 samples (5 samples per condition). A random subset of genes is chosen to be differentially expressed, with the number determined by the specified proportion of differential expression. 

\textbf{Asymmetry, same mRNA/cell:} Differential expression is asymmetric (more genes up-regulated under one condition than the other), but the absolute expression is the same for each condition. 75$\%$ of DE genes were 2-fold up-regulated under condition A, and 25$\%$ were 4-fold up-regulated under condition B.

\textbf{Asymmetry, different mRNA/cell:} Differential expression is asymmetric, and the absolute expression is different under the different conditions. 75$\%$ of DE genes were 2-fold up-regulated under condition A, and 25$\%$ were 2-fold up-regulated under condition B.

\textbf{Symmetry, same mRNA/cell:} Differential expression is symmetric (same number up-regulated under each condition), and the absolute expression is the same for each condition. 50$\%$ of DE genes were 2-fold up-regulated under condition A, and 50$\%$ were 2-fold up-regulated under condition B.

\textbf{Symmetry, different mRNA/cell:} Differential expression is symmetric, but the absolute expression is different under the different conditions. 50$\%$ of genes are 4-fold up-regulated under condition A, and 50$\%$ are 6-fold up-regulated under condition B.

\section*{EVALUATION OF METHODS AND ASSUMPTIONS}

\begin{table}
\begin{small}
\begin{center}
\begin{tabular}{|p{15mm}|p{90mm}|p{40mm}|}
\hline
Paper Goal & \centering Evaluation Criteria & Approximate Ranking\\
\hline
Global compare & Equiv. normalized count distribution between replicates (real data); variance of normalized counts within condition (real data); equiv. expression of housekeeping genes (real data); agreement on DE calls (real data); false positives and power (simulation). \cite{dillies2013comprehensive} & DESeq $\&$ TMM \newline UQ $\&$ Med \newline Q \newline RPKM $\&$ TC  \\
\hline
Introduces UQ & DE detection compared to qRT-PCR (ROC curves) (real data); variability between replicates after normalization (real data); bias in fold-change estimation compared to qRT-PCR (real data). \cite{bullard2010evaluation} & UQ \newline Q \newline TC \\
\hline
Introduces MRN & False positives, false negatives, and power (simulation); MSE of expression fold-change estimates (simulation); number of DE calls and agreement on DE calls (real data). \cite{maza2013comparison} & MRN \newline DESeq $\&$ TMM \newline TC \newline UQ $\&$ Med \newline FPKM \\
\hline
Global compare & Equiv. normalized count distribution between replicates (real data); variance of normalized counts within condition (real data); agreement on DE calls (real data); variability of results under different filtering techniques (real data). \cite{lin2016comparison} & DESeq \newline TMM \newline UQ, Med, $\&$ Q \newline RPKM $\&$ TC \newline (RUVg considered, but assumptions not met)\\
\hline
Global compare & Correlation between normalized counts and qRT-PCR data (real and simulated data). \cite{li2015comparing} & All were equivalent \newline (DESeq, Med, Q, RPKM and ERPKM, TMM, UQ) \\
\hline
Global compare & Bias and variance in fold change estimation (compared to housekeeping genes) (real data); sensitivity and specificity in DE calls (using genes believed to be DE and non-DE) (real data); prediction of DE genes (real data); agreement on DE calls (real data). \cite{zyprych2015impact}  & DESeq \newline PS \newline Q \newline UQ \newline TMM\\
\hline
Global compare & Clustering of normalized counts agrees with condition (real data); correlation between fold change estimates and qRT-PCR fold changes (real data). \cite{rapaport2013comprehensive} & All were equivalent \newline (DESeq, PS, UQ, TMM, Q, CuffDiff) \\
\hline
Introduces DEGES & ROC curves and AUC (real and simulated data). \cite{kadota2012normalization} & DEGES strategy using a normalization method generally performed better than that method by itself \\
\hline
Introduces CLS & Observed fold change for normalized data (real data). \cite{loven2012revisiting} & CLS \newline RPKM\\
\hline
Introduces RUV & PCA (real data); variance and distribution of normalized data (real data); distribution of p-values (real data); clustering and proportion of reads mapping to spike-ins (real data); MA plots (real data); ROC curves (real data); comparison with qRT-PCR (real data). \cite{risso2014normalization} & RUV \newline (UQ, CLS, RPKM, TMM, DESeq, \& Q)\\
\hline

\end{tabular}
\end{center}
\caption{\small \textbf{Literature comparing normalization methods.} Several papers which include comparisons of DE assumption normalization methods are summarized here. Short descriptions of the criteria used to evaluate the normalization methods are provided, and the final results of the paper are condensed into an approximate ranking of the methods considered (best performing methods at the top). These rankings are not explicit in all papers and for some have been inferred from the paper's discussion of the strengths and weaknesses of the different methods. Abbreviations: UQ = Upper Quartile, Med = Median, Q = Quantile, TC = Total Count, MRN = Median Ratio, PS = PoissonSeq, CLS = Cyclic Loess on Spike-ins.}
\label{tab: litSummary}
\end{small}
\end{table}

Several papers have investigated the different normalization methods described in the previous section. Table \ref{tab: litSummary} summarizes these comparisons by giving approximate rankings of the methods evaluated in each comparison. Here we expand on these rankings to remark on several key ideas.

\textbf{Differences in mRNA/cell result in poor performance of library size normalization.} As shown in Table \ref{tab: litSummary}, in many comparisons Total Count and RPKM/FPKM perform worse than all other methods, and several authors expressly recommend against its use \cite{dillies2013comprehensive}. A likely cause of this is that in these evaluations, the assumption required for library size normalization (same amount of mRNA/cell) is violated. For example, Dillies \textit{et al.} \cite{dillies2013comprehensive} observed that a few highly expressed genes had a large share of the read counts in \textit{M. musculus} data they compared. Bullard \textit{et al.} \cite{bullard2010evaluation} and Lin \textit{et al.} \cite{lin2016comparison} reported similar findings. Bullard saw 50$\%$ of the reads concentrated in 5$\%$ of the genes, and Lin found 50$\%$ of the reads in 45 genes for male flies and 186 genes for female flies. With such a large proportion of the reads aligned to a small fraction of the genes, if these genes are differentially expressed it is likely that there will be different amounts of mRNA/cell across the conditions, and Bullard \textit{et al.} \cite{bullard2010evaluation} did observe that the highly expressed genes were differentially expressed.

\textbf{DESeq and TMM generally perform well, but validity is not certain.} Dillies \textit{et al.} \cite{dillies2013comprehensive}, for example, note that DESeq and TMM are the only methods that perform well both with the ability to detect differentially expressed genes, and with controlling false positives. This supports the conclusion of Bullard \textit{et al.} \cite{bullard2010evaluation}, who concluded that normalization has the biggest impact on detection of differentially expressed genes.

Given that several authors have found that a few highly expressed genes have a large share of total expression \cite{dillies2013comprehensive, bullard2010evaluation, lin2016comparison} and these genes may be differentially expressed, it is clear that assuming the same amount of mRNA/cell is not always reasonable. The good performance of DESeq and TMM in these studies suggests that perhaps their assumptions (DE and non-DE genes behave the same, balanced expression) are fairly reasonable, or at least not too violated, for the data analyzed in the comparisons. However, it is possible that for the real data analyzed in these comparisons there is a global shift in expression that is not picked up by these normalization methods. For example, a global shift has been observed in DE analysis with low and high c-Myc conditions \cite{lin2012transcriptional, nie2012cMyc}, and this shift was undetected without the use of spike-in controls \cite{loven2012revisiting}. Other researchers \cite{athanasiadou2016growth, hu2014nucleosome} have found similar global up-regulation when using spike-ins, and it has been suggested that such shifts were not detected by previous research due to lack of proper normalization \cite{hu2014nucleosome}. Even qRT-PCR, often treated as a ``gold standard" for evaluating the performance of DE analysis methods, might not be able to detect a global shift without controls. Normalization for qRT-PCR often relies on housekeeping genes \cite{bullard2010evaluation, lee2002control, kanno2006percell}. In the absence of non-DE genes, as occurs with a global shift in expression, qRT-PCR results might not be accurate. Furthermore, the use of PCR as a gold standard for evaluation of normalization methods has been called into question, as despite being highly accurate, PCR can contain errors \cite{sun2012systematic}. Hence, for methods which normalize by distribution or by testing, it is difficult or impossible to know whether their assumptions have been met without additional information. 

\textbf{Potential lack of housekeeping genes.} The possible absence of housekeeping genes poses a problem for HG normalization of RNA-Seq data as well as PCR data. While Bullard \textit{et al.} \cite{bullard2010evaluation} found that HG normalization performed equivalently to UQ, the housekeeping gene they used (POLR2A) was selected based on previous studies and they caution that such information may not always be available. Dillies \textit{et al.} \cite{dillies2013comprehensive} also selected housekeeping genes from previous research, and state that one cannot be certain housekeeping genes will always be non-DE. As mentioned above, several authors have found global shifts in expression which would leave few, if any, non-DE housekeeping genes for use in normalization \cite{athanasiadou2016growth, hu2014nucleosome, lin2012transcriptional, nie2012cMyc}.

\textbf{External controls may be needed.} In the case of a global shift in expression, the assumptions are violated for normalization methods that do not rely on external controls. Global up-regulation necessarily leads to different amounts of mRNA/cell (library size normalization), highly asymmetric expression (distribution/testing normalization), and an absence of non-DE genes (HG normalization).  Without the use of external controls, it is possible that many experiments have resulted in incorrect conclusions \cite{chen2015overlooked}. Normalization with spike-in controls attempts to rectify the problems of asymmetry, by relying on genes/spike-ins which should have the same expression under the different conditions. 

\textbf{Mixed performance of spike-ins.} As we have seen, these methods come with their own set of assumptions, and it is not clear that these assumptions can always be trusted. In an assessment of ERCC spike-in controls, Jiang \textit{et al.} found that only small fractions ($0.5\%$ and $0.01\%$) of spike-in reads were incorrectly aligned to the actual genome of the organisms in their experiment (\textit{Drosophila} and humans) \cite{jiang2011synthetic}. This indicates that as desired, there will be little error introduced into the read counts by the controls. Furthermore, Jiang \textit{et al.} found a linear relationship between the amount of spike-in and read count \cite{jiang2011synthetic}, which is evidence that the spike-in read counts are representative of expression level. However, Risso \textit{et al.} \cite{risso2014normalization} found violations of both assumptions necessary for basic spike-in normalization (spike-ins are non-DE across conditions and have the same technical effects as genes), and Qing \textit{et al.} \cite{qing2013mRNA} found that read counts for the spike-ins depended in part on the mRNA enrichment protocol used in the experiment.

\textbf{Recommendations: appropriate method depends on DE definition and assumptions.} Different circumstances call for different normalization methods. Correct normalization should cause non-DE genes to have the same (expected) normalized read count across conditions. This requires a definition of differential expression. In this paper, we defined differential expression in terms of differences in mRNA/cell across conditions, and it appears that this is the definition used in previous research evaluating normalization methods. Consequently, the majority of the commentary and recommendations presented here is in the context of mRNA/cell differential expression.  However, other definitions of differential expression are possible and may be appropriate/necessary in certain conditions \cite{coate2015variation}. One alternative is to define a gene as differentially expressed if its share of mRNA in the transcriptome is different across conditions; this bases differential expression on \textit{relative}, rather than \textit{absolute}, measures of expression. The mRNA/transcriptome definition may be appropriate in some circumstances: Ignatov \textit{et al.} \cite{ignatov2015dormant} performed an experiment which found down-regulation of every gene when using the mRNA/cell definition, so they chose instead to look for differences in per transcriptome expression.

Choosing a normalization method depends on the definition of differential expression. For example, library size normalization generally performs poorly when defining DE in terms of mRNA/cell, but should produce exactly the desired measure when defining DE in terms of mRNA/transcriptome. Hence, choosing a normalization method for an RNA-Seq experiment must begin with choosing a definition of differential expression.

Once differential expression is defined, the next step is to determine which assumptions are appropriate for the experiment at hand, and then choose a method that follows those assumptions. Assumptions of each method depend on the definition of differential expression; in this paper, we consider the assumptions necessary for each method under mRNA/cell differential expression. However, these assumptions will not be the same for mRNA/transcriptome differential expression. For example, the assumption for library size normalization discussed above is that the total mRNA/cell is the same under each condition. This assumption is necessary for the relative measures of expression obtained via library size normalization to be valid measures of absolute expression. If a relative definition of DE is used instead, such as mRNA/transcriptome, then it is not necessary to assume equivalent total mRNA/cell across conditions.

If spike-ins can be trusted, they are important to use in normalization because there may be previously unknown shifts in expression that cannot be detected without controls, and housekeeping genes do not seem a reliable choice for controls. RUV aims to address the shortcomings of spike-ins, so may be a good method to use when spike-ins are available.

However, there are situations in which spike-in methods are not an option. Coate and Doyle \cite{coate2015variation} note that application of spike-in methods requires the ability to count the number of cells used in RNA extraction, and cell counting is not possible in some tissue types. In these cases, normalization by distribution/testing appears to be the best option, and DESeq especially has generally been shown to perform well.

\section*{CONCLUSION}
The use of RNA-Seq experiments to study organisms' genomes is becoming ubiquitous, and the explosion in the use of sequencing technology has led to a related explosion in the development of statistical methods for processing and analyzing RNA-Seq data. As previous research has demonstrated \cite{bullard2010evaluation}, proper normalization is an essential step in the analysis pipeline. We have seen that incorrect normalization can result in downstream errors such as inflated false positives. The need for normalization arises from the inherent variability in the collection of RNA-Seq data, and a variety of normalization methods have been devised to combat this variability. As we have seen, the literature has not reached a consensus on which normalization method to use. 

Each normalization procedure relies on assumptions, and when violated the procedures lead to incorrect results. For each assumption, there is evidence that it may not hold in some experiments. Part of an analysis of RNA-Seq data requires choosing a normalization procedure, and keeping the assumptions of each method in mind can help to make the appropriate choice for the experiment at hand. However, there may be many situations in which the validity of any assumption is unknown for the given experiment. In such cases, normalization with external controls would be the appropriate choice if the external controls can be trusted. Unfortunately, several authors have found problems with spike-ins and so propose additional methods to handle these issues. It is clear that spike-ins are necessary in some circumstances, and we hope that as research progresses their performance will improve. 

To the best of our knowledge, there does not exist an extensive analysis of published data which evaluates the assumptions of normalization methods.  Given the potential violations to each normalization assumption, knowledge of the extent to which each assumption holds in a given experiment would be instrumental in helping to choose a normalization method for RNA-Seq analysis. There is no clear way to perform such an evaluation, however, considering that violations of assumptions (such as a global shift) may go undetected without additional information and the requisite information may not be present in the original experiment.

\section*{SUPPLEMENTARY INFORMATION}

\subsection*{The False Discovery Rate}
Following the notation of Benjamini and Hochberg \cite{benjamini1995controlling}, suppose there is a family of $m$ independent hypotheses to be tested, $m_0$ of which are truly null. We represent the different possible outcomes in Table \ref{tab: bhTable}. 

\begin{table}[H]
\begin{small}
\begin{center}
\begin{tabular}{|c c c c|}
\hline
$ $ & {\it Declared} & {\it Declared } & {\it Total}\\
$ $ & {\it non-significant} & {\it significant} & $ $\\
\hline
True null hypotheses & {\bf U} & {\bf V} & $m_0$\\
Non-true null hypotheses & {\bf T} & {\bf S} & $m - m_0$\\
$ $ & $m - ${\bf R} & {\bf R} & $m$\\
\hline
\end{tabular}
\end{center}
\end{small}
\caption{\small{True discoveries and false discoveries when testing $m$ null hypotheses \cite{benjamini1995controlling}.}}
\label{tab: bhTable}
\end{table}

\begin{defn}{\emph{\textbf{The false discovery rate.}}}
\label{defn: FDR}
\emph{From Table \ref{tab: bhTable}, the proportion of discoveries which are false is $V/R$, and the {\bf false discovery rate} is defined to be
 \begin{align*}
 FDR = E(V/R)
 \end{align*}
 where $V/R = 0$ whenever $R = 0$. In other words, 
 \begin{align*}
 FDR = E(V/R | R >0)P(R > 0).
 \end{align*}}
\end{defn}
 
To control the FDR at a desired level $\alpha$, Benjamini and Hochberg proposed the following step-up procedure (henceforth referred to as BH) \cite{benjamini1995controlling}. 

\begin{defn}{\emph{\textbf{BH procedure.}}}
\label{defn: BHProc}
\emph{Let $p_1,...,p_m$ be the p-values resulting from tests of the $m$ hypotheses, and $p_{(1)}, p_{(2)},...,p_{(m)}$ the p-values in increasing order. The {\bf BH procedure} finds the largest index $i$ such that
\begin{align*}
p_{(i)} \leq \alpha \frac{i}{m}
\end{align*}
and then $p_{(1)},...,p_{(i)}$ are declared significant, and their associated hypotheses rejected. Equivalently, each p-value $p_{(i)}$ is adjusted by setting $p_{(i)} = \min\{\frac{m}{j}p_{(j)} : j \geq i\}$, then all p-values below the cutoff $\alpha$ are rejected. }

\end{defn}

Benjamini and Hochberg proved that this procedure controls the FDR at 
\begin{align*}
FDR \leq \alpha \frac{m_0}{m} \leq \alpha
\end{align*}
and furthermore that the cutoff $T = \max\{p_{(i)} : p_{(i)} \leq \alpha \frac{i}{m}\}$ can be less stringent than the cutoff given by FWER control, since FWER control implies FDR control but a procedure controlling the FDR need not necessarily control the FWER \cite{benjamini1995controlling}.

In the two decades since the introduction of the FDR, a number of alternative approaches have been suggested, including related errors like Storey's positive false discovery rate (pFDR) \cite{storey2003pfdr}, and adaptive methods for controlling the FDR while attempting to maximize power, such as the one proposed by Storey and Tibshirani \cite{storey2003pnas}. Other methods and procedures attempt to control FDR in more complicated scenarios. %\cite{Reiner2006}. 
The common goal of all these varied methods is to maintain error control in different situations while conserving as much power as possible.

While more advanced methods than the BH procedure are demonstrably better at controlling FDR, in the sense of maintaining control while increasing power (the method proposed in \cite{storey2003pnas} is one such example) the most common choice appears to still be BH, and is in fact the default in the several DE packages. For this reason, FDR control performed in simulations in this paper is done using BH.

\subsection*{Details on Normalization}
Here we provide more specifics on the normalization procedures mentioned in the body of the text.\\

{\bf Total Count:} Total count normalization deals with the most observable difference in RNA-Seq samples: their library sizes. In total count normalization \cite{dillies2013comprehensive}, read counts are normalized by dividing each count by the total number of reads in its sample. The goal of total count normalization is to account for differences in library size by simply dividing by library size in each sample.\\

{\bf RPKM:} RPKM (reads per kilobase per million mapped reads) normalization \cite{mortazavi2008mapping} is an adaptation of total count normalization that attempts to normalize by gene length as well as the total number of reads in each sample. As the name suggests, in RPKM normalization each read count is normalized by dividing by the number of reads in the sample (in millions) and the gene length (in kilobases).\\

{\bf FPKM:} FPKM (fragments per kilobase per million mapped fragments) normalization \cite{trapnell2010transcript} is almost exactly the same as RPKM normalization, with the change of using cDNA molecules rather than RNA reads; each cDNA molecule corresponds to two reads, each starting at a different end of the fragment.\\

{\bf Quantile:} Before the use of RNA-Seq experiments was common, a huge body of work was developed for the analysis of microarray data.  Quantile normalization is the result of applying a normalization method used in microarray analysis to RNA-Seq data.  The basic algorithm is as follows, and is designed to make use of the fact that data vectors with the same distribution will have their quantiles plotted on the diagonal, by forcing the normalized data to have quantiles on the diagonal and hence have the same distribution \cite{bolstad2003comparison}:
\begin{enumerate}
\item Sort each column of the read count matrix; this causes each row to contain the same quantiles of each sample.
\item Replace each entry in the sorted read count matrix with the mean of that row.
\item Undo the sorting on the read count matrix, so that the entries are now back in the original order.
\end{enumerate}
Using this algorithm, the read count matrix has been normalized so that each sample is forced to have the same distribution over all the genes. Other measures such as the median could be used in place of the mean of the quantiles.\\

{\bf Upper Quartile:} Upper quartile normalization \cite{bullard2010evaluation} is similar to quantile normalization but focuses on one specific quantile (the 75$^{\text{th}}$ percentile). In upper quartile normalization, each read count is divided by the 75$^{\text{th}}$ percentile of the read counts in its sample, where genes with read counts of 0 across all samples are excluded. Zyprych-Walczak \textit{et al.} \cite{zyprych2015impact} also report a variant of Upper Quartile normalization in a rather complicated form that ultimately reduces to scaling each Upper Quartile normalization factor by the geometric means of the Upper Quartiles, so that the product of the normalization factors is 1.\\

{\bf Median:} Median normalization \cite{dillies2013comprehensive} is essentially the same as Upper Quartile normalization, except that gene counts are scaled by the median of counts in their sample rather than the 75$^{\text{th}}$ percentile.\\

{\bf DESeq:} The DESeq normalization strategy attempts to find a \textit{size factor} for each sample, such that the ratios of size factors of different samples represent the ratio of their respective sequencing depths. Let $k_{ij}$ be the number of reads aligned to gene $i$ under sample $j$. The estimated size factor $\hat{s}_j$ for sample $j$ is given by

\begin{align*}
\hat{s}_j = \text{median}_i \left\lbrace \frac{k_{ij}}{\left( \prod \limits_{v=1}^m k_{iv} \right)^{1/m}} \right\rbrace
\end{align*}
where $m$ is the total number of samples, across all conditions. The denominator $\left( \prod \limits_{v=1}^m k_{iv} \right)^{1/m}$ serves as a pseudo-reference sample to which each sample can be compared.  As discussed in \cite{anders2010differential}, the rational behind the size factor estimation is that a good estimate for the ratio of sequencing depths of two samples should be the median of the ratios of their counts.  This is generalized to multiple samples through the use of the pseudo-reference sample.\\

{\bf CuffDiff:} Introduced by Trapnell \textit{et al.} \cite{trapnell2013differential} as part of the CuffDiff 2 software, the CuffDiff normalization method is a slight modification of the DESeq method. The CuffDiff approach calculates two different normalization factors: an \textit{internal scale} is used when comparing samples taken under the same biological conditions, while an \textit{external scale} is used to compare samples across different biological conditions. 

Calculation of the internal scale is simply a restriction of the DESeq normalization method to the read count sub-matrix for each set of replicates; in an experiment with two conditions $A$ and $B$ and three replicates per condition, for example, the DESeq method would be applied to both groups of replicates separately, taking three columns of the matrix with each application.

The external scale is calculated after the internal scale; in the case of 5 samples per condition and two conditions, the result would be 10 size factors. Let $\hat{s}_j$ denote the internal size factor for sample $j$. We then use the internal size factors to normalize each column (divide by the corresponding internal size factor). For each gene and each condition, we average the internal-scaled counts for the replicates in that gene and condition; let $\overline{k}_{i,A}$ and $\overline{k}_{i, B}$ denote these averages for gene $i$ in the case of two conditions. That is, with $k_{ij}$ again denoting the $(i,j)$ entry of the full read count matrix,
\begin{align*}
\overline{k}_{i,A} = \frac{1}{m_A} \sum \limits_{j: \rho(j) = A} \frac{k_{ij}}{\hat{s}_j}
\end{align*}
and likewise for $\overline{k}_{i,B}$, where $m_A$ is the number of samples performed under condition $A$ and $\rho(j)$ denotes the condition under which sample $j$ was performed. We then use the $\overline{k}_{i,\rho(j)}$ values to produce external size factor estimates 
\begin{align*}
\eta_j = \text{median}_i \left \lbrace \overline{k}_{i,\rho(j)}\left( \prod \limits_{\rho(v)} \overline{k}_{i,\rho(v)} \right)^{-1/c} \right \rbrace
\end{align*}
where $c$ is the number of conditions.  To compare internal-scaled counts across different conditions (such as for DE testing), we adjust the internal-scaled counts using the external scale.\\

{\bf TMM:} TMM (Trimmed Mean of the M-values) \cite{robinson2010scaling} is a normalization strategy with a very similar approach to the DESeq size-factor estimate.  TMM sets one of the samples as a reference sample, then compares the counts in each sample to the reference sample to estimate the ratio of sequencing depths between each sample and the reference.  The procedure involves trimming genes twice, using both the fold-changes and expression levels between samples; the goal is to remove genes that are differentially expressed, so that the mean can be taken over genes that do not show differential expression.  For these genes, we expect that the ratio of counts in one sample to the reference sample is represented by the ratio of the sequencing depths.

Let $k_{ij}$ again denote the number of reads aligned to gene $i$ under sample $j$. Let $\mu_{ij}$ be the true gene expression level of gene $i$ under sample $j$, and $N_j$ the total number of reads for sample $j$, i.e. the library size $\left( N_j = \sum \limits_{i} k_{ij} \right)$. Fixing one of the samples $r$ as the reference sample, we define \textit{gene-wise log fold changes}
\begin{align*}
M_{ij}^r = \log_2 \frac{k_{ij}/N_j}{k_{ir}/N_r}
\end{align*}
and \textit{absolute expression levels}
\begin{align*}
A_{ij}^r = \frac{1}{2} \log_2 \left( \frac{k_{ij}}{N_j} \cdot \frac{k_{ir}}{N_r} \right).
\end{align*}
For sample $j$, the $M_{ij}^r$ and $A_{ij}^r$ values are trimmed independently (the default amount trimmed is 30$\%$ for the $M_{ij}^r$ and $5\%$ for the $A_{ij}^r$) to produce a set of genes $G$ for which neither the $M_{ij}^r$ nor $A_{ij}^r$ value was removed (trimmed). Using this set $G$, we calculate the scaling factor $TMM_j^{(r)}$ for sample $j$ via a weighted mean:
\begin{align*}
\log_2(TMM_j^{(r)}) = \frac{\sum \limits_{i \in G} w_{ij}^r M_{ij}^r}{\sum \limits_{i \in G} w_{ij}^r}
\end{align*}
where
\begin{align*}
w_{ij}^r = \frac{N_j - k_{ij}}{N_jk_{ij}} - \frac{N_r - k_{ir}}{N_rk_{ir}}.
\end{align*}

Note that in the calculation of the scaling factors, we divide by the library size of each sample $(N_j)$. Thus, the $TMM_j^{(r)}$ scaling factors tell us the relative size of samples after we have normalized by library size, and to normalize so that read counts are directly comparable between samples we would divide each sample by $TMM_j^{(r)} \cdot \dfrac{N_j}{N_r}$ where $N_r$ is the library size of the reference sample.\\

{\bf Median Ratio:} Similarly to how CuffDiff normalization extends the DESeq normalization procedure, Median Ratio normalization (MRN) \cite{maza2013comparison} is designed to be a more robust adaptation of the TMM method.  As in the TMM method, define $k_{ij}$ to be the number of reads aligned to gene $i$ under sample $j$ and $N_j$ the number of reads in sample $j$ (its library size). And like the TMM method, the MRN method separates library size normalization and normalization of the samples after dividing by library size. Here, as in \cite{maza2013comparison}, we will describe MRN in the special case where there are two experimental conditions $A$ and $B$, although the method can be generalized to more than two conditions.

MRN begins by taking the mean of library-normalized counts for each gene within each condition:
\begin{align*}
\overline{k}_{iA} = \frac{1}{m_A} \sum \limits_{j : \rho(j) = A} \frac{k_{ij}}{N_j}
\end{align*}
would define this mean for condition $A$, and the definition is analogous for condition $B$. Then, we calculate the ratio $\tau_i$ of these two means for each gene $i$:
\begin{align*}
\tau_i = \frac{\overline{k}_{iB}}{\overline{k}_{iA}}.
\end{align*}
We define $\tau$ to be the median of these ratios across all genes.  The intuition is that between two samples of the same experimental condition, the difference in sequencing depth can be determined directly by the difference in library size since there are no genes which can be differentially expressed within the same biological condition.  Then, normalization by library size puts samples within the same condition on the same scale.  Any remaining differences in normalized read counts within a replicate group are then due to randomness, and so we can remove some of that natural variability by averaging across samples within a replicate group. Then, $\tau$ represents the median relative size of samples under each condition after accounting for library size; to get the normalization factor for the original read count matrix, we include the library size:
\begin{align*}
e_j = \begin{cases} 
      N_j & \text{ if } \rho(j) = A \\
      \tau \cdot N_j & \text{ if } \rho(j) = B  
   \end{cases}
\end{align*}
Then, dividing each column of the original read count matrix by its corresponding normalization factor will allow for direct comparison of reads across different samples and conditions.  The final step is to make the product of the normalization factors be 1 by dividing by their geometric mean, which does not change the relationship between them but ensures that the normalized read counts will be on a similar scale as the originals. Let $\tilde{f} = \left( \prod \limits_{v=1}^m e_v \right)^{1/m}$ where $m$ is the total number of samples across all conditions.  Then, the final normalization factor for sample $j$ is
\begin{align*}
f_j = \frac{e_j}{\tilde{f}}
\end{align*}
\\

{\bf PoissonSeq:} The information for normalization is found in the non-differentially expressed genes. TMM explicitly aims to remove differentially expressed genes through trimmed means, leaving the non-DE genes as the set of genes used in estimates. Methods like Upper Quartile normalization, DESeq, and MRN address the issue by examining a quartile of the data, or a transformed version of the data, that is expected to be reasonably representative of the non-differentially expressed genes. In the PoissonSeq method \cite{li2012normalization}, developed as part of the PoissonSeq package, the idea of using the non-differentially expressed genes is taken a step further by directly performing a goodness-of-fit test to try to find a subset of non-differentially expressed genes.

Let $K_{ij}$ be the random variable for the number of reads aligned to gene $i$ under sample $j$. It is assumed in the PoissonSeq package that $K_{ij} \sim \text{Poisson}(\mu_{ij})$, although for the purposes of the normalization technique the most salient point is using $\mu_{ij}$ to denote the expectation of $K_{ij}$, and the actual distribution of $K_{ij}$ is less important for normalization than for performing tests for differential expression. We model $\mu_{ij}$ using
\begin{align*}
\log(\mu_{ij}) = \log(d_j) + \log(\beta_i) + \gamma_{i,\rho(j)}
\end{align*}
where $d_j$ is the sequencing depth for sample $j$, $\beta_i$ is the level of expression of gene $i$, and $\gamma_{i, \rho(j)}$ represents how associated the expression of gene $i$ is with the condition $\rho(j)$ of sample $j$.  If $\gamma_i$ is 0 for all conditions, then there is no association between the expression of gene $i$ and the biological conditions and hence gene $i$ is not differentially expressed in the study. Under the null hypothesis that there is no association between gene $i$ and the condition of sample $j$, $\gamma_{i,\rho(j)} = 0$.

We estimate the expression level of gene $i$ as $\hat{\beta}_i = \sum \limits_{v=1}^m k_{iv}$ where $m$ is the total number of samples across all conditions. Since sequencing depth can be compared across samples using non-differentially expressed genes, given a set $S$ of non-differentially expressed genes we can compute an estimate for the sequencing depth of sample $j$ by the proportion of reads aligned to non-differentially expressed genes that come from sample $j$:
\begin{align*}
\hat{d}_j = \frac{\sum \limits_{i \in S} k_{ij}}{\sum \limits_{i \in S} \left( \sum \limits_{v=1}^m k_{iv} \right)} = \frac{\sum \limits_{i \in S} k_{ij}}{\sum \limits_{i \in S} \hat{\beta}_i}.
\end{align*}

For genes in $S$, $\gamma_{i,\rho(j)} = 0$ and so $\log(\mu_{ij}) = \log(d_j \beta_i)$. Hence, an estimate for $E(k_{ij})$ is $\hat{d}_j\hat{\beta}_i$ and we can create a goodness-of-fit statistic for each gene $i$:
\begin{align*}
GOF_i = \sum \limits_{v=1}^m \frac{(k_{ij} - \hat{d}_j \hat{\beta}_i)^2}{\hat{d}_j \hat{\beta}_i}.
\end{align*}
We ultimately want a good estimate of $d_j$, which means we want to identify $S$. To do so, we start with an initial estimate of $d_j$ using the entire set of genes as $S$, then calculate $GOF_i$ statistics and take the middle $(1 - 2\varepsilon) \cdot 100\%$ and re-calculate $\hat{d}_j$.  We then alternate between estimating $S$ and $d_j$ until convergence.  By default, PoissonSeq uses $\varepsilon = 0.25$. The final sequencing depth estimates $\hat{d}_j$ are then scaled so that their product is 1.
\\

{\bf DEGES:} This normalization approach \cite{kadota2012normalization}, which stands for Differentially Expressed Gene Elimination Strategy, has a very similar approach to PoissonSeq. It alternates between estimating normalization factors and using those normalization factors to determine which genes are differentially expressed.  We will describe the algorithm without relying on a specific strategy for normalization or testing.
\begin{enumerate}
\item Using all genes in the experiment, calculate normalization factors for each sample.  For example, if we used DESeq normalization, we would calculate the median of the relative expression values across all genes.
\item Using the normalization factors from Step 1, perform differential expression hypothesis testing and identify a set of non-differentially expressed genes.
\item Re-calculate normalization factors using the set of genes identified in Step 2.
\end{enumerate}
The algorithm alternates between Steps 2 and 3 a prespecified number of times, the idea being to iteratively improve normalization. The final normalization factors can then be used in an official differential expression analysis.
\\

{\bf Housekeeping Genes:} If one can identify \textit{a priori} a set of non-DE genes, these could be used for normalization purposes. For example, Bullard \textit{et al.} \cite{bullard2010evaluation} investigates the use of housekeeping genes, specifically POLR2A, to perform normalization. With one gene, all read counts in a sample are scaled by a single factor, so that after normalizing each sample the read counts for the housekeeping gene are the same across all samples. With multiple housekeeping genes, typical normalization methods can be applied to the set of housekeeping genes rather than to all genes. For example, DESeq normalization could be applied to the read count matrix restricted to the housekeeping genes, then the size factor estimates obtained would be applied to normalize the entire read count matrix as usual.
\\

{\bf Spike-in Controls:} As with housekeeping genes, typical normalization methods like Upper Quartile and DESeq can be applied to only the spike-in controls \cite{risso2014normalization}, producing normalization factors that are then applied to all genes. Another spike-in method was proposed by Lov{\'e}n \textit{et al.}, who used loess normalization to normalize the RPKM values for all genes to the spike-in RPKM values \cite{loven2012revisiting}. As in Lov{\'e}n \textit{et al.} \cite{loven2012revisiting}, spike-in normalization typically requires that the spikes be added in proportion to the number of cells from which the sample RNA is extracted \cite{coate2015variation}; this ensures that the spikes will have the same RNA/cell in each condition. Changes in the proportion of reads aligned to the spikes in a sample then indicate changes in the amount of mRNA/cell for the genes, which can be reflected in the read counts by adjusting counts to equilibrate the spike-in counts across samples (see Figure \ref{fig: simpleNegControl}).

Loess normalization, originally developed to normalize microarray intensities, can also be applied to normalize RPKM values. The method works as follows \cite{bolstad2003comparison, loven2012revisiting}, and compares two samples at a time. First, M and A values for each gene/spike-in $i$ are calculated between samples $j$ and $k$ (similar to the M-A values calculated in TMM):
\begin{align*}
M_i &= \log_2 \left( \frac{RPKM_{ij}}{RPKM_{ik}} \right)\\
A_i &= \frac{1}{2} \log_2 \left( RPKM_{ij} \cdot RPKM_{ik} \right)
\end{align*}
Consider the $M_i$ values for the spike-ins. Since the spike-ins should be non-DE, then we would expect each $M_i$ value to be 0. Thus, we want to adjust the $M_i$ values so that when we plot adjusted $M_i$ against $A_i$, the adjusted $M_i$ values are scattered around 0. To do so, plot the $M_i$ against $A_i$ for the spikes and fit a loess curve to the data (note that the loess curve is fit only with the spike-ins); let $\hat{M}_i$ be the fitted value on the curve for each gene/spike. Then, the adjusted $M_i$ value is
\begin{align*}
M_i' = M_i - \hat{M}_i.
\end{align*}
All genes, not just the spike-ins, are adjusted in this way. Since the loess curve was calculated using only the spike-ins, $M_i'$ for the spike-ins will be centered around 0 as desired, but we avoid centering all genes around 0 and so are still able to detect shifts in expression.

To calculate the re-normalized values, $RPKM_{ij}'$ and $RPKM_{ik}'$,
\begin{align*}
RPKM_{ij}' &= 2^{A_i + \frac{M_i'}{2}}\\
RPKM_{ik}' &= 2^{A_i - \frac{M_i'}{2}}.
\end{align*}
Re-normalized RPKM values are calculated for each pair of samples, then the original RPKM values are corrected using each of the pairwise corrections.
\\

{\bf Remove Unwanted Variation:} Adapted from previous work on normalization of microarray data, the Remove Unwanted Variation \cite{risso2014normalization} (RUV) method aims to remove variation between samples that is not the result of the biological covariates of interest. The notation associated with this method will differ from that used in the other normalization procedures described above, as the method is sufficiently complicated that it is easiest to communicate by being consistent with the original paper.

Suppose an RNA-Seq experiment is performed with $J$ genes and $n$ samples, and $p$ covariates of interest. We will restrict our examination of this method to the classic case of a differential study with two conditions.  In this case, $p = 2$.
\begin{itemize}
\item Let $Y \in \mathsf{M}_{n \times J}$ be the read count matrix, so $Y_{ij}$ corresponds to the number of reads aligned to gene $j$ in sample $i$.
\item Let $X \in \mathsf{M}_{n \times p}$ denote the design matrix for the experiment. In our restricted case, the design matrix has a column for the intercept and each entry in the second column is an indicator for whether the sample corresponding to that row is under condition $A$ or condition $B$.
\item Let $W \in \mathsf{M}_{n \times k}$ be a matrix related to $k$ factors of unwanted variance ($k$ must be specified beforehand).
\item Let $\alpha \in \mathsf{M}_{k \times J}$ be the coefficients corresponding to the factors of unwanted variance in $W$.
\item Let $\beta \in \mathsf{M}_{p \times J}$ be the coefficients which represent the relationship between each gene and each covariate of interest.
\item Let $O \in \mathsf{M}_{n \times J}$ be a matrix reflecting sequencing depth offsets; the authors suggest using Upper Quartile normalization, though of course other methods would also work in its place.
\end{itemize}
Then, we assume the log-linear model
\begin{align}
\log E[Y | W, X, O] = W\alpha + X\beta + O.
\label{eqn: RUVmodel}
\end{align}

The RUV method provides three different sub-procedures to approach normalization given this model, with varying assumptions.  RUVg uses that a set of negative control genes (which can be spike-in controls) is known. RUVr uses the residuals of a first-pass fit to the log-linear model in Equation (\ref{eqn: RUVmodel}) and does not require knowledge of negative control genes, though does assume that the factors of unwanted variation are uncorrelated with the biological conditions. RUVs creates negative control samples by comparing samples within replicate groups, and also relies on negative control genes and the factors of unwanted variation being uncorrelated with the biological conditions in the experiment. The difference between RUVs and RUVg is that RUVs is designed to be more robust to the choice of negative control genes, and the authors state that the method can still perform reasonably even when the entire set of genes is used.

The three RUV paths are reasonably similar, and so for sake of brevity only one (RUVg) will be described here; notation is borrowed from Risso \textit{et al.} \cite{risso2014normalization}. We begin by assuming that there is a set of $J_c$ negative control genes. When the matrices in Equation (\ref{eqn: RUVmodel}) are restricted to the negative control genes, we will use the subscript $c$.
\begin{enumerate}
\item Define $Z_c = \log Y_c - O_c$, so that we have accounted for offsets in the experimental data. This should make samples of different sequencing depths comparable. Then let $Z^*_c$ be the column-centered version of $Z_c$. After accounting for sequencing depth, the only variation of negative control genes across samples is from factors of unwanted variation. By subtracting the mean of each column, the measurement of the expression of each gene in $Z^*_c$ is centered at 0, which also allows the intercept term to be 0 in $\beta_c$. Since none of the genes are associated with the biological covariates of interest, the other coefficients in $\beta_c$ will be 0 as well, yielding $Z^*_c = W \alpha_c$.
\item Next, perform the singular value decomposition of $Z^*_c$, so $Z^*_c = U \Lambda V^T$ where $\Lambda$ is the rectangular diagonal matrix of singular values of $Z^*_c$. 
\item For a given number $k$ of factors of unwanted variation, we are interested in determining the impact of those factors so we reduce to only the $k$ largest singular values. Denote by $\Lambda_k$ the $n \times J_c$ matrix obtained from $\Lambda$ by setting all singular values but the $k$ largest to 0. We estimate $W$ by $\hat{W} = U \Lambda_k$ where we have removed columns of 0s to ensure that $\hat{W} \in \mathsf{M}_{n\times k}$. Under the assumption that the factors of unwanted variation for the negative control genes span the same space as the factors of unwanted variation for all genes (in the linear algebra sense, since columns of $W$ are factors of unwanted variation and $W\alpha$ is a linear combination of the columns of $W$), then $\hat{W}$ will estimate $W$.
\item Substituting $\hat{W}$ back into Equation (\ref{eqn: RUVmodel}), and with knowledge of the design matrix $X$, GLM regression can be used to estimate the remaining parameters $\alpha$ and $\beta$, and then differential expression analysis can be performed.  Though the authors do not recommend obtaining normalized counts separately from the differential expression analysis procedure, it is possible to use RUVg to normalize by performing OLS regression of $Z = \log Y - O$ on $\hat{W}$.  The residuals of this regression are the normalized read counts.
\end{enumerate}

We also present some intuition to further explain RUVg. We don't care about $W$ or $\alpha$ separately, as the normalization considers only their product $W\alpha$. As $W\alpha$ is a linear combination of the columns of $W$, then $W\alpha$ could be represented infinitely many ways by replacing the columns of $W$ with another set of vectors spanning the same space and replacing $\alpha$ by the correct coefficients to get the same linear combination with the new spanning set. Hence, if we assume that the factors of unwanted variation for the negative control genes span the same space as the factors of unwanted variation for all genes, then $W$ can be calculated using only the negative control genes since we just need to span the same space rather than get exactly the same matrix.

\section*{KEY POINTS}
\begin{itemize}
\item Assumptions allow normalization to translate raw read counts into meaningful measures of expression.
\item The correct normalization method to use depends on which assumptions are valid for the biological experiment.
\item Incorrect normalization leads to problems in downstream analysis, such as inflated false positives, that mean results cannot be trusted.
\item There are examples of global shifts in expression that violate assumptions of conventional normalization methods, requiring controls.
\item No normalization method is perfect, and for every method there exists cases for which the assumptions are violated.
\item An understanding of assumptions can help pick the most suitable normalization method for a given experiment.
\end{itemize}

\section*{FUNDING}
This research was supported in part by grants to Pomona College [52007555] and Harvey Mudd College [52007544] from the Howard Hughes Medical Institute through the Precollege and Undergraduate Science Education Program.

\bibliographystyle{brief}
\bibliography{research_citations}

\begin{thebibliography}{10}

\bibitem{shendure2008beginning}
Shendure J.
\newblock The beginning of the end for microarrays?
\newblock \emph{Nat Methods} 2008;\hspace{0pt}\textbf{5}(7):585--7.

\bibitem{oshlack2010rna}
Oshlack A, Robinson MD, Young MD.
\newblock From {RNA-seq} reads to differential expression results.
\newblock \emph{Genome Biol} 2010;\hspace{0pt}\textbf{11}(12):1--10.

\bibitem{wang2009rna}
Wang Z, Gerstein M, Snyder M.
\newblock {RNA-Seq}: a revolutionary tool for transcriptomics.
\newblock \emph{Nat Rev Genet} 2009;\hspace{0pt}\textbf{10}(1):57--63.

\bibitem{auer2011differential}
Auer PL, Srivastava S, Doerge R.
\newblock Differential expression - the next generation and beyond.
\newblock \emph{Brief Funct Genomics} 2012;\hspace{0pt}\textbf{11}(1):57--62.

\bibitem{oshlack2009length}
Oshlack A, Wakefield MJ.
\newblock Transcript length bias in {RNA-seq} data confounds systems biology.
\newblock \emph{Biol Direct} 2009;\hspace{0pt}\textbf{4}(1):1--10.

\bibitem{risso2011gc}
Risso D, Schwartz K, Sherlock G, \emph{et~al.}
\newblock {GC}-content normalization for {RNA-Seq} data.
\newblock \emph{BMC Bioinformatics} 2011;\hspace{0pt}\textbf{12}(1):1--17.

\bibitem{robinson2010scaling}
Robinson MD, Oshlack A, \emph{et~al.}
\newblock A scaling normalization method for differential expression analysis
  of {RNA-seq} data.
\newblock \emph{Genome Biol} 2010;\hspace{0pt}\textbf{11}(3):1--9.

\bibitem{mcintyre2011technical}
McIntyre L, Lopiano K, Morse A, \emph{et~al.}
\newblock {RNA-seq}: technical variability and sampling.
\newblock \emph{BMC Genomics} 2011;\hspace{0pt}\textbf{12}(1):1--13.

\bibitem{dillies2013comprehensive}
Dillies MA, Rau A, Aubert J, \emph{et~al.}
\newblock A comprehensive evaluation of normalization methods for {I}llumina
  high-throughput {RNA} sequencing data analysis.
\newblock \emph{Brief Bioinform} 2013;\hspace{0pt}\textbf{14}(6):671--83.

\bibitem{bullard2010evaluation}
Bullard JH, Purdom E, Hansen KD, \emph{et~al.}
\newblock Evaluation of statistical methods for normalization and differential
  expression in {mRNA-Seq} experiments.
\newblock \emph{BMC Bioinformatics} 2010;\hspace{0pt}\textbf{11}(1):1--13.

\bibitem{kadota2012normalization}
Kadota K, Nishiyama T, Shimizu K.
\newblock A normalization strategy for comparing tag count data.
\newblock \emph{Algorithms Mol Biol} 2012;\hspace{0pt}\textbf{7}(1):1--13.

\bibitem{li2015comparing}
Li P, Piao Y, Shon H, \emph{et~al.}
\newblock Comparing the normalization methods for the differential analysis of
  {I}llumina high-throughput {RNA-Seq} data.
\newblock \emph{BMC Bioinformatics} 2015;\hspace{0pt}\textbf{16}(1):1--9.

\bibitem{lin2016comparison}
Lin Y, Golovnina K, Chen Z, \emph{et~al.}
\newblock Comparison of normalization and differential expression analyses
  using {RNA-Seq} data from 726 individual \emph{Drosophila melanogaster}.
\newblock \emph{BMC Genomics} 2016;\hspace{0pt}\textbf{17}(1):1--20.

\bibitem{maza2013comparison}
Maza E, Frasse P, Senin P, \emph{et~al.}
\newblock Comparison of normalization methods for differential gene expression
  analysis in {RNA-Seq} experiments: {A} matter of relative size of studied
  transcriptomes.
\newblock \emph{Commun Integr Biol} 2013;\hspace{0pt}\textbf{6}(6).

\bibitem{rapaport2013comprehensive}
Rapaport F, Khanin R, Liang Y, \emph{et~al.}
\newblock Comprehensive evaluation of differential gene expression analysis
  methods for {RNA-seq} data.
\newblock \emph{Genome Biol} 2013;\hspace{0pt}\textbf{14}(9):1--13.

\bibitem{zyprych2015impact}
Zyprych-Walczak J, Szabelska A, Handschuh L, \emph{et~al.}
\newblock The impact of normalization methods on {RNA-seq} data analysis.
\newblock \emph{BioMed Res Int} 2015;\hspace{0pt}\textbf{2015}:621690.

\bibitem{athanasiadou2016growth}
Athanasiadou N, Neymotin B, Brandt N, \emph{et~al.}
\newblock Growth rate-dependent global amplification of gene expression.
\newblock \emph{bioRxiv} 2016;\hspace{0pt}.

\bibitem{hu2014nucleosome}
Hu Z, Chen K, Xia Z, \emph{et~al.}
\newblock Nucleosome loss leads to global transcriptional up-regulation and
  genomic instability during yeast aging.
\newblock \emph{Genes Dev} 2014;\hspace{0pt}\textbf{28}(4):396--408.

\bibitem{lin2012transcriptional}
Lin C, Lov\'{e}n J, Rahl P, \emph{et~al.}
\newblock Transcriptional amplification in tumor cells with elevated c-{M}yc.
\newblock \emph{Cell} 2012;\hspace{0pt}\textbf{151}(1):56--67.

\bibitem{nie2012cMyc}
Nie Z, Hu G, Cui K, \emph{et~al.}
\newblock c-{M}yc is a universal amplifier of expressed genes in lymphocytes
  and embryonic stem cells.
\newblock \emph{Cell} 2012;\hspace{0pt}\textbf{151}(1):68--79.

\bibitem{chen2015overlooked}
Chen K, Hu Z, Xia Z, \emph{et~al.}
\newblock The overlooked fact: fundamental need for spike-in controls for
  virtually all genome-wide analyses.
\newblock \emph{Mol Cell Biol} 2015;\hspace{0pt}\textbf{36}(5):662--7.

\bibitem{pachter2011models}
Pachter L.
\newblock Models for transcript quantification from {RNA-Seq}.
\newblock \emph{arXiv} 2011;\hspace{0pt}.

\bibitem{loven2012revisiting}
Lov\'{e}n J, Orlando D, Sigova A, \emph{et~al.}
\newblock Revisiting global gene expression analysis.
\newblock \emph{Cell} 2012;\hspace{0pt}\textbf{151}(3):476--82.

\bibitem{coate2015variation}
Coate JE, Doyle JJ.
\newblock Variation in transcriptome size: are we getting the message?
\newblock \emph{Chromosoma} 2015;\hspace{0pt}\textbf{124}(1):27--43.

\bibitem{anders2010differential}
Anders S, Huber W.
\newblock Differential expression analysis for sequence count data.
\newblock \emph{Genome Biol} 2010;\hspace{0pt}\textbf{11}(10):1--12.

\bibitem{mortazavi2008mapping}
Mortazavi A, Williams BA, McCue K, \emph{et~al.}
\newblock Mapping and quantifying mammalian transcriptomes by {RNA-Seq}.
\newblock \emph{Nat Methods} 2008;\hspace{0pt}\textbf{5}(7):621--8.

\bibitem{trapnell2010transcript}
Trapnell C, Williams BA, Pertea G, \emph{et~al.}
\newblock Transcript assembly and quantification by {RNA-Seq} reveals
  unannotated transcripts and isofrom switching during cell differentiation.
\newblock \emph{Nat Biotechnol} 2010;\hspace{0pt}\textbf{28}(5):511--5.

\bibitem{bolstad2003comparison}
Bolstad B, Irizarry R, {\AA}strand M, \emph{et~al.}
\newblock A comparison of normalization methods for high density
  oligonucleotide array data based on variance and bias.
\newblock \emph{Bioinformatics} 2003;\hspace{0pt}\textbf{19}(2):185--93.

\bibitem{robinson2010edgeR}
Robinson MD, McCarthy DJ, Smyth GK.
\newblock edge{R}: a {B}ioconductor package for differential expression
  analysis of digital gene expression data.
\newblock \emph{Bioinformatics} 2010;\hspace{0pt}\textbf{26}(1):139--40.

\bibitem{li2012normalization}
Li J, Witten D, Johnstone I, \emph{et~al.}
\newblock Normalization, testing, and false discovery rate estimation for
  {RNA-}sequencing data.
\newblock \emph{Biostatistics} 2012;\hspace{0pt}\textbf{13}(3):523 -- 38.

\bibitem{eisenberg2013housekeeping}
Eisenberg E, Levanon E.
\newblock Human housekeeping genes, revisted.
\newblock \emph{Hum Genet} 2013;\hspace{0pt}\textbf{29}(10):569--74.

\bibitem{jiang2011synthetic}
Jiang L, Schlesinger F, Davis C, \emph{et~al.}
\newblock Synthetic spike-in standards for {RNA-seq} experiments.
\newblock \emph{Genome Res} 2011;\hspace{0pt}\textbf{21}(9):1543--51.

\bibitem{risso2014normalization}
Risso D, Ngai J, Speed T, \emph{et~al.}
\newblock Normalization of {RNA-seq} data using factor analysis of control
  genes or samples.
\newblock \emph{Nat Biotechnol} 2014;\hspace{0pt}\textbf{32}(9):896--902.

\bibitem{law2014voom}
Law CW, Chen Y, Shi W, \emph{et~al.}
\newblock {v}oom: precision weights unlock linear model analysis tools for
  {RNA-seq} read counts.
\newblock \emph{Genome Biol} 2014;\hspace{0pt}\textbf{15}(2):1--17.

\bibitem{love2014moderated}
Love MI, Huber W, Anders S.
\newblock Moderated estimation of fold change and dispersion for {RNA-seq} data
  with {DESeq2}.
\newblock \emph{Genome Biol} 2014;\hspace{0pt}\textbf{15}(12):1--21.

\bibitem{benjamini1995controlling}
Benjamini Y, Hochberg Y.
\newblock Controlling the false discovery rate: a practical and powerful
  approach to multiple testing.
\newblock \emph{J R Stat Soc Series B Stat Methodol}
  1995;\hspace{0pt}\textbf{57}(1):289--300.

\bibitem{lee2002control}
Lee P, Sladek R, Greenwood C, \emph{et~al.}
\newblock Control genes and variability: absence of ubiquitous reference
  transcripts in diverse mammalian expression studies.
\newblock \emph{Genome Res} 2002;\hspace{0pt}\textbf{12}(2):292--7.

\bibitem{kanno2006percell}
Kanno J, Aisaki K, Igarashi K, \emph{et~al.}
\newblock ``{P}er cell" normalization method for {mRNA} measurement by
  quantitative {PCR} and microarrays.
\newblock \emph{BMC Genomics} 2006;\hspace{0pt}\textbf{7}(1):1--14.

\bibitem{sun2012systematic}
Sun Z, Zhu Y.
\newblock Systematic comparison of {RNA-Seq} normalization methods using
  measurement error models.
\newblock \emph{Bioinformatics} 2012;\hspace{0pt}\textbf{28}(20):2584--91.

\bibitem{qing2013mRNA}
Qing T, Yu Y, Du T, \emph{et~al.}
\newblock {mRNA} enrichment protocols determine the quantification
  characteristics of external {RNA} spike-in controls in {RNA-Seq} studies.
\newblock \emph{Sci China Life Sci} 2013;\hspace{0pt}\textbf{56}(2):134--42.

\bibitem{ignatov2015dormant}
Ignatov DV, Salina EG, Fursov MV, \emph{et~al.}
\newblock Dormant non-culturable \emph{Mycobacterium tuberculosis} retains
  stable low-abundant {mRNA}.
\newblock \emph{BMC Genomics} 2015;\hspace{0pt}\textbf{16}(1):1--13.

\bibitem{storey2003pfdr}
Storey JD.
\newblock The positive false discovery rate: a {B}ayesian interpretation and
  the q-value.
\newblock \emph{Ann Stat} 2003;\hspace{0pt}\textbf{31}(6):2013--35.

\bibitem{storey2003pnas}
Storey JD, Tibshirani R.
\newblock Statistical significance for genomewide studies.
\newblock \emph{Proc Natl Acad Sci USA}
  2003;\hspace{0pt}\textbf{100}(16):9440--5.

\bibitem{trapnell2013differential}
Trapnell C, Hendrickson G, Sauvageau M, \emph{et~al.}
\newblock Differential analysis of gene regulation at transcript resolution
  with {RNA-Seq}.
\newblock \emph{Nat Biotechnol} 2013;\hspace{0pt}\textbf{31}:46--53.

\end{thebibliography}

\end{document}